\documentclass[12pt,journal,final,onecolumn]{IEEEtran}

\usepackage[dvips]{graphicx}
\usepackage{epsfig}
\usepackage[cmex10]{amsmath}
\usepackage{amssymb}
\usepackage{amsthm}
\usepackage{amsfonts}
\usepackage{bm}
\usepackage{cite}
\usepackage[svgnames]{xcolor} 
\usepackage{pstricks,pst-node,pst-plot,pstricks-add}
\usepackage[tight,footnotesize]{subfigure}
\usepackage{algpseudocode}
\usepackage{algorithm}
\usepackage{psfrag, enumerate}

\IEEEoverridecommandlockouts

\input{niesen.def}

\newcommand{\rf}{r}

\begin{document}

\title{Hierarchical Coded Caching}
\author{Nikhil Karamchandani, Urs Niesen, Mohammad Ali Maddah-Ali, and Suhas Diggavi%
    \thanks{N. Karamchandani and S. Diggavi are with UCLA, U. Niesen and
        M. A. Maddah-Ali are with Bell Labs, Alcatel-Lucent.
        Emails: nikhil@ee.ucla.edu, urs.niesen@alcatel-lucent.com,
        mohammadali.maddah-ali@alcatel-lucent.com, suhasdiggavi@ucla.edu
    }%
}

\maketitle

\begin{abstract} 
    Caching of popular content during off-peak hours is a strategy to     
    reduce network loads during peak hours. Recent work has shown
    significant benefits of designing such caching strategies not only
    to deliver part of the content locally, but also to provide coded
    multicasting opportunities even among users with different demands.
    Exploiting both of these gains was shown to be approximately optimal
    for caching systems with a single layer of caches. 

    Motivated by practical scenarios, we consider in this work a
    hierarchical content delivery network with two layers of caches. We
    propose a new caching scheme that combines two basic approaches.
    The first approach provides coded multicasting opportunities within
    each layer; the second approach provides coded multicasting
    opportunities across multiple layers. By striking the right balance
    between these two approaches, we show that the proposed scheme
    achieves the optimal communication rates to within a constant
    multiplicative and additive gap. We further show that there is no
    tension between the rates in each of the two layers up to the
    aforementioned gap. Thus, both layers can simultaneously operate at
    approximately the minimum rate.
\end{abstract}

\section{Introduction}
\label{Sec:Intro}

The demand for high-definition video streaming services such as YouTube
and Netflix is driving the rapid growth of Internet traffic. In
order to mitigate the effect of this increased load on the underlying
communication infrastructure, content delivery networks deploy storage
memories or caches throughout the network. These caches can be populated
with some of the  content during off-peak traffic hours. This
cached content can then be used to reduce the network load during peak
traffic hours when users make the most requests. 

Content caching has a rich history, see for example~\cite{Wessels:2001}
and references therein. More recently, it has been studied in the
context of video-on-demand systems for which efficient content placement
schemes have been proposed in~\cite{Borst:2010, Tan:2013} among others.
The impact of different content popularities on the performance of caching schemes has
been investigated for example in~\cite{Wolman99, breslau99,
Applegate:2010}. A common feature among the caching schemes studied in
the literature is that those parts of a requested file that are
available at nearby caches are served locally, whereas the remaining
files parts are served via orthogonal transmissions from an origin
server hosting all the files. 

Recently, \cite{CachingUM, DCachingUM} proposed a new caching approach,
called \emph{coded caching}, that exploits cache memories not only to
deliver part of the content locally, but also to create coded
multicasting opportunities among users with different demands. It is
shown there that the reduction in rate due to these coded multicasting
opportunities is significant and can be on the order of the number of
users in the network.  The setting considered in \cite{CachingUM,
DCachingUM} consists of a single layer of caches between the origin
server and the end users. The server communicates directly with all the
caches via a shared link, and the objective is to minimize the required
transmission rate by the server. For this basic network scenario, coded
caching is shown there to be optimal within a constant factor.  These
results have been extended to nonuniform demands in~\cite{niesen13, multilevel14} and nonuniform cache-access in \cite{multiaccess14}, and to online caching systems in~\cite{pedarsani13}. 

In practice, many caching systems consist of not only one but multiple
layers of caches, usually arranged in a tree-like hierarchy with the
origin server at the root node and the users connected to the leaf
caches~\cite{danzig1996, korupolu99, Borst:2010}. Each parent cache
communicates with its children caches in the next layer, and the
objective is to minimize the transmission rates in the various layers.

There are several key questions when analyzing such hierarchical caching
systems. A first question is to characterize the optimal tradeoff
between the cache memory sizes and the rates of the links connecting the
layers. One particular point of interest is if there is any
tension between the rates in the different layers in the network. In
other words, if we reduce the rate in one layer, does it necessarily
increase the rate in other layers? If there is no such tension, then
both layers can simultaneously operate at minimum rate. A second
question is how to extend the coded caching approach to this setting.
Can we can simply apply the single-layer scheme from \cite{CachingUM,
DCachingUM} in each layer separately or do we need to apply coding
across several layers in order to minimize transmission rates? 

\begin{figure}[ht]
    \centerline{\includegraphics{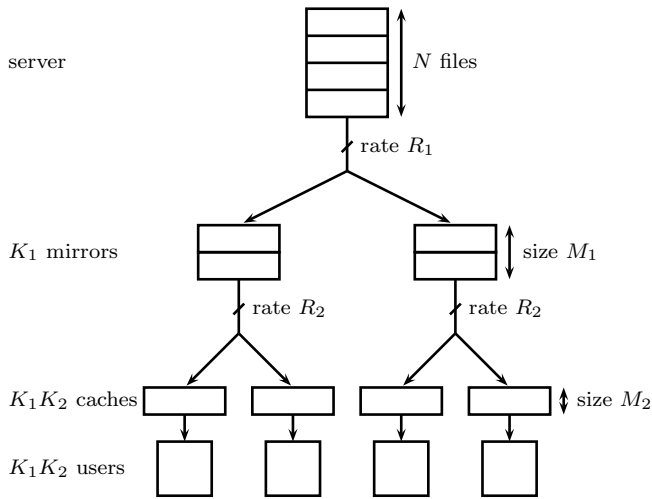}}
    \caption{ System setup for the hierarchical caching problem: A server
        hosting $N$ files is connected to $K_1$ mirrors each able to
        store $M_1$ of the files. Each of the mirrors, in turn, is
        connected to $K_2$ caches each able to store $M_2$ of the files.
        A single user is attached to each of these caches. Once the
        mirrors and caches are filled, each user requests one of the $N$
        files. The aim is to minimize the rate $R_1$ from the server to
        the mirrors and the rate $R_2$ from the mirrors to the caches.
        In the figure, $N=4$, $K_1=K_2=2$, $M_1=2$, and $M_2=1$. }
    \label{Fig:Setup}
\end{figure}

In this work, we focus on a hierarchical caching system with two layers
of caches as depicted in Fig.~\ref{Fig:Setup}. For simplicity, we will
refer to the first layer of caches as mirrors.  We propose a new caching
scheme exploiting two types of coded caching opportunities: The first
type involves only a single layer at a time, i.e., it operates between a
node and its direct children.  These single-layer coding opportunities
are available over the link connecting the origin server to the mirrors
and also in the link connecting each mirror to the user caches. The
second type involves two layers at a time. These two-layer opportunities
are available between the origin server and the user caches. We show
that, by striking the right balance between these two types of coded
caching opportunities, the proposed caching scheme attains the
approximately optimal memory-rate tradeoff to within a constant additive
and multiplicative gap. Due to the possible interaction between the two
cache layers, the network admits many different prefetching and delivery
approaches. It is thus perhaps surprising that a combination of these
two basic schemes is sufficient to achieve the approximately optimal
memory-rate tradeoff. Furthermore, investigating the achievable rates
also reveals that there is no tension between the rates over the first
and second layers up to the same aforementioned gap. Thus, both layers
can simultaneously operate at approximately minimum rate.

The remainder of the paper is organized as follows. We describe the
problem setting in Section~\ref{Sec:ProblemSetup} and provide some
preliminaries in Section~\ref{Sec:SingleLayer}.
Section~\ref{Sec:Results} presents our main results and discusses their
engineering implications.  Section~\ref{Sec:AchievableSchemes}
introduces the proposed caching scheme and characterizes its
performance. The proofs of our main results are discussed in
Section~\ref{Sec:Proofs} and their details are provided in the
appendices. Appendix~\ref{Sec:LowerBounds} proves information-theoretic
bounds on the performance of any caching scheme. The proof of the
constant multiplicative and additive gap between the performance of the
proposed scheme and the optimal caching scheme is provided in
Appendices~\ref{Sec:GapR1} and \ref{Sec:GapR2}.

\section{Problem Setting}
\label{Sec:ProblemSetup}

We consider a hierarchical content delivery network as illustrated in
Fig.~\ref{Fig:Setup} in Section~\ref{Sec:Intro}. The system consists of
a single origin server hosting a collection of $N$ files each of size
$F$ bits. The server is connected through an error-free broadcast link
to $K_1$ mirror sites, each with memory of size $M_{1}F$ bits. Each
mirror, in turn, is connected through an error-free broadcast link to
$K_2$ users. Thus, the system has a total of $K_1K_2$ users.  Each user
has an associated cache memory of size $M_{2}F$ bits. The quantities
$M_1$ and $M_2$ are the normalized memory sizes of the mirrors and user
caches, respectively.  We refer to the $j$th user attached to mirror $i$
as ``user $(i,j)$'' and the corresponding cache as ``cache ($i,j$)''.
Throughout, we will focus on the most relevant case where the number of
files $N$ is larger than the total number of users $K_1K_2$ in the
system\footnote{For example, in a video application such as Netflix, each ``file" corresponds to a short segment of a video, perhaps a few seconds to a minute long. If there are $1000$ different popular movies of length $100$ minutes each, this would correspond to more than $100,000$ different files.}, i.e., $N \ge K_1K_2$. 

The content delivery system operates in two phases: a \emph{placement
phase} followed by a \emph{delivery phase}. The placement phase occurs
during a period of low network traffic. In this phase, all the mirrors
and user caches store content related to the $N$ files (possibly using
randomized strategies), while satisfying the corresponding memory
constraints. Crucially, this is done without any prior knowledge of
future user requests. The delivery phase occurs during a period of high
network traffic. In this phase, each user requests one of the $N$ files
from the server. Formally, the user requests can be represented as a
matrix $\bm{D}$ with entry $d_{i,j}\in\{1,2,\dots,N\}$ denoting the
request of user $(i,j)$.  The user requests are forwarded to the
corresponding mirrors and further on to the server. Based on the
requests and the stored contents of the mirrors and the user caches
during the placement phase, the server transmits a message $X^{\bm{D}}$
of size at most $R_{1}F$ bits over the broadcast link to the mirrors.
Each mirror $i$ receives the server message and, using its own memory
content, transmits a message $Y^{\bm{D}}_{i}$ of size at most $R_{2}F$
bits over its broadcast link to users $(i,1), (i,2), \ldots, (i,K_2)$.
Using only the contents of its cache $(i,j)$ and the received message
$Y^{\bm{D}}_{i}$ from mirror $i$, each user $(i, j)$ attempts to
reconstruct its requested file $d_{i,j}$. 

For a given request matrix $\bm{D}$, we say that the tuple $\left(M_1,
M_2, R_1, R_2\right)$ is \emph{feasible for request matrix $\bm{D}$}
if, for large enough file size $F$, each user $(i,j)$ is able to recover
its requested file $d_{i, j}$ with probability\footnote{The
feasibility of a tuple corresponds to a random variable because of the
possible randomization of the placement and delivery phases.}
arbitrarily close to one. We say that the tuple $\left(M_1, M_2, R_1,
R_2\right)$ is \emph{feasible} if it is feasible for all possible
request matrices $\bm{D}$. The object of interest in the remainder of
this paper is the feasible rate region:

\begin{definition}
    \label{Defn:FeasibleRegion}
    For memory sizes $M_1, M_2 \geq 0$, the \emph{feasible rate region}
    is defined as
    \begin{equation}
        \label{Eqn:FeasibleRegion}
        \mc{R}^\star(M_1, M_2) 
        \defeq \cl \bigl\{ 
            (R_1, R_2) : \text{$(M_1, M_2, R_1, R_2)$ is feasible}  
        \bigr\}.
    \end{equation}
\end{definition}

\section{Preliminaries}
\label{Sec:SingleLayer}

The proposed achievable scheme for the hierarchical caching setting
makes use of the coded caching scheme developed for networks with a
single layer of caches. In this section, we recall this single-layer
caching scheme.

Consider the special case of the hierarchical caching setting with no
cache memory at the users and only a single user accessing each mirror,
i.e., $M_2 = 0$ and $K_2 = 1$. Let the normalized mirror memory size be
$M_1 = M$ and the number of mirrors $K_1 = K$.  This results in a system
with only a single layer of caches (namely the mirrors).

Note that for this single-layer scenario, each mirror is forced to recover
the file requested by its corresponding user and then forward the
entire file to it. Thus, a transmission rate of $R_2 = K_2 = 1$ over the
link from the mirror to the user is both necessary and sufficient in
this case. The goal is to minimize the transmission rate $R_1$ from the
server to the mirrors. 

This single-layer setting was recently studied in \cite{CachingUM,
DCachingUM}, where the authors proposed a coded caching scheme. For
future reference, we recall this scheme in Algorithm~\ref{alg:1} and
illustrate it below in Example~\ref{Ex:Basic}. The authors showed that
rate $R_1 = \rf(M / N, K)$ is feasible in this setting,  where
$\rf(\cdot, \cdot)$ is given by 
\begin{equation}
    \label{Eqn:BasicRate} 
    \rf\!\left(\frac{M}{N}, K\right) 
    \defeq \left[K\cdot \left( 1 - \frac{M}{N} \right) 
    \cdot \frac{N}{KM}\left( 1 - \left(1 - \frac{M}{N}\right)^K\right) \right]^+
\end{equation}
with 
$$
[x]^+ \defeq \max\{x, 0\}.
$$
The right hand side
of~\eqref{Eqn:BasicRate} consists of three terms. The first term is the
rate without caching. The second term, called \emph{local caching gain},
represents the savings due to a fraction of each file being locally
available. The third term, called \emph{global caching gain}, is the
gain due to coding. It is shown in~\cite{DCachingUM} that this
achievable rate $R_1$ is within a constant factor of the minimum
achievable rate for this single-layer setting for any value of $N$, $K$,
and $M$.  We will refer to the placement and delivery procedures of the
single-layer coded caching scheme in Algorithm~\ref{alg:1} as
BasePlacement($N, K, M$) and BaseDelivery($N, K, M$), respectively. 

\begin{algorithm}
    \caption{Single-Layer Coded Caching~\cite{DCachingUM}}
    \label{alg:1}
    \begin{algorithmic}[1]
        \Statex \hspace{-\labelwidth}\textbullet\hspace{3pt} 
        $[K] \defeq \{1,2,\ldots, K\}, [N] \defeq \{1,2,\ldots, N\}$ 
        \Statex \hspace{-\labelwidth}\textbullet\hspace{3pt}
        Request vector $\bm{d} = \left(d_1, d_2, \ldots, d_{K} \right)$
        \Statex \hspace{-\labelwidth}\textbullet\hspace{3pt}
        In Line~\ref{alg:1_send}, $\oplus$ denotes bit-wise XOR operation. For any
        subset $\mc{S} \subset [K]$ of mirrors,  $V_{j, \mc{S}}$ denotes the bits of
        file $d_j$ requested by user $j$ stored exclusively at mirrors in $\mc{S}$. 
        \Statex 
        \Procedure{BasePlacement}{}
        \For{$i\in[K], n\in[N]$}
        \State mirror $i$ independently stores a subset of  $\tfrac{MF}{N}$ bits  of file $n$, chosen uniformly at random 
        \label{alg:1_cache}
        \EndFor
        \EndProcedure
        \Statex
        \Procedure{BaseDelivery}{$\bm{d}$}
        \For{$s=K, K-1, \ldots, 1$} \label{alg:1_sloop}
        \For{$\mc{S}\subset[K]: \card{\mc{S}}=s$} \label{alg:1_Sloop}
        \State server sends \(\oplus_{j\in\mc{S}} V_{j,\mc{S}\setminus\{j\}}\) \label{alg:1_send}
        \EndFor 
        \EndFor 
        \EndProcedure
    \end{algorithmic}
\end{algorithm}

\begin{example}[\emph{Single-Layer Coded Caching~\cite{DCachingUM}}]
    \label{Ex:Basic}
    Consider the single-layer setting as described above with $N=2$
    files and $K = 2$ mirrors each of size $M \in [0,2]$. For ease
    of notation, denote the files by $A$ and $B$. In the
    placement phase of Algorithm~\ref{alg:1}, each mirror stores a
    subset of $MF / N = MF / 2$ bits of each of the two files, chosen
    uniformly and independently at random. Each bit of a file is thus
    stored in a given mirror with probability $M / N = M / 2$. 

    Consider file $A$ and notice that we can view it as being composed
    of $2^{K} = 4$ subfiles 
    \begin{equation*}
        A = \left( A_{\emptyset}, A_{1}, A_{2}, A_{1,2}\right), 
    \end{equation*}
    where $A_{\mc{S}}$ denotes the bits of file $A$ which are
    exclusively stored in the mirrors in $\mc{S}$. For example, $A_{1}$
    denotes the bits of file $A$ which are stored only in mirror one,
    and $A_{1,2}$ denotes the bits of file $A$ which are available in
    both mirrors one and two. For large enough file size $F$, we have by
    the law of large numbers that  
    \begin{equation*}
        \card{ A_{\mc{S} } } \approx 
        \left(\frac{M}{2}\right)^{ \card{\mc{S} } } 
        \left( 1 - \frac{M}{2} \right)^{2 - \card{\mc{S} }} F 
    \end{equation*}
    for any subset $\mc{S}$. File $B$ can similarly be partitioned into subfiles. 

    In the delivery phase, suppose for example that the first user
    requests file $A$ and the second user requests file $B$.
    By Line~\ref{alg:1_send} in Algorithm~\ref{alg:1}, the server
    transmits $A_2 \oplus B_1$, $A_{\emptyset}$, and $B_{\emptyset}$
    where $\oplus$ denotes bit-wise XOR. 
    
    Consider mirror one whose corresponding user has requested file $A$.
    Mirror one already knows the subfiles $A_1, A_{1,2}$ from its cache
    memory.  Further, the server's transmission provides the subfile
    $A_{\emptyset}$. Finally, from $A_{2} \oplus B_{1}$ transmitted by the
    server, the mirror can recover $A_{2}$ since it has $B_1$ stored in
    its cache memory. Thus, from the contents of its memory and the
    server transmission, mirror one can recover $A = \left( A_{\emptyset},
    A_{1}, A_{2}, A_{1,2}, \right)$ and then forward it to its attached
    user. Similarly, mirror two can recover file $B$ and forward it to
    its attached user. The number of bits transmitted by the server is
    given by 
    \begin{equation*}
        \frac{M}{2} \left( 1 - \frac{M}{2} \right)F + 2\left( 1 - \frac{M}{2} \right)^{2} F = 2 \cdot \left(1 - \frac{M}{2} \right) \cdot \frac{2}{2M}\left( 1 - \left( 1 - \frac{M}{2} \right)^2 \right) F.  
    \end{equation*}
    which agrees with the expression in \eqref{Eqn:BasicRate}.
\end{example}

While the above discussion focuses on $K_2 = 1$ user accessing each
mirror, the achievable scheme can easily be extended to $K_2 > 1$ 
by performing the delivery phase in $K_2$ stages with one unique
user per mirror active in each stage. From~\cite[Section V]{DCachingUM}, the resulting rate over the
first link is
\begin{equation}
    \label{Eqn:BasicRateK2} 
    R_1 = K_2 \cdot \rf(M_1 / N, K_1).
\end{equation}

\section{Main Results}
\label{Sec:Results}

As the main result of this paper, we provide an approximation of the
feasible rate region $\mc{R}^\star(M_1, M_2)$ for the general
hierarchical caching problem with two layers. We start by introducing
some notation.  For $\alpha, \beta \in [0,1]$, define the rates
\begin{subequations}
    \label{Eqn:TotalRates}
    \begin{align}
        \label{Eqn:TotalRatesa}
        R_{1}(\alpha, \beta) 
        & \defeq  \alpha K_2 \cdot \rf\!\left(\frac{M_1}{\alpha N}, K_1\right) 
        + (1-\alpha)\cdot \rf\!\left(\frac{(1-\beta)M_2}{(1-\alpha)N}, K_1K_2 \right), \\ 
        \label{Eqn:TotalRatesb}
        R_{2}(\alpha, \beta) 
        & \defeq \alpha \cdot \rf\!\left(\frac{\beta M_2}{\alpha N}, K_2 \right) 
        + (1-\alpha) \cdot \rf\!\left(\frac{(1-\beta)M_2}{(1-\alpha)N},
        K_2 \right),
    \end{align}
\end{subequations}
where $\rf(\cdot, \cdot)$ is defined in~\eqref{Eqn:BasicRate} in
Section~\ref{Sec:SingleLayer}. Next, consider the following region: 
\begin{definition}
    \label{Defn:AchievableRatesG}
    For memory sizes $M_1, M_2 \geq 0$, define 
    \begin{equation}
        \label{Eqn:AchievableRatesG}
        \mc{R}_C( M_1, M_2) 
        \defeq  \bigl\{ \bigl( R_1(\alpha, \beta), R_2(\alpha, \beta) \bigr) 
        : \alpha, \beta \in [0, 1] \bigr\}  \; + \;
        \mathbb{R}^2_+,
    \end{equation}
    where $\mathbb{R}^2_+$ denotes the positive quadrant, $R_1(\alpha,
    \beta), R_2(\alpha, \beta)$ are defined in \eqref{Eqn:TotalRates},
    and the addition corresponds to the Minkowski sum between sets.  
\end{definition}
As will be discussed in more detail later, the set
$\mc{R}_C(M_1, M_2)$ is the rate region achieved by appropriately
sharing the available memory between two basic achievable
schemes during the placement phase and then using each scheme to recover
a certain fraction of the requested files during the delivery phase.
Each of these two schemes is responsible for one of the two terms in
$R_1(\alpha, \beta)$ and $R_2(\alpha, \beta)$. The parameters $\alpha$
and $\beta$ dictate  what fraction of
each file and what fraction of the memory is allocated to each of these two schemes.  The set
$\mc{R}_C(M_1, M_2)$ is thus the rate region achieved by all
possible choices of the parameters $\alpha$ and $\beta$.

Our main result shows that, for any memory sizes $M_1, M_2$, the region
$\mc{R}_C(M_1, M_2)$ just defined approximates the
feasible rate region $\mc{R}^\star(M_1, M_2)$. 

\begin{theorem}
    \label{Theorem:Bounds}
    Consider the hierarchical caching problem in Fig.~\ref{Fig:Setup}
    with $N$ files, $K_1$ mirrors, and $K_2$ users accessing each
    mirror. Each mirror and user cache has a normalized memory size of
    $M_1$ and $M_2$, respectively. Then we have 
    \begin{equation*}
        \mc{R}_C(M_1, M_2) 
        \subseteq \mc{R}^\star(M_1, M_2) 
        \subseteq  c_1 \cdot \mc{R}_C(M_1, M_2) - c_2,
    \end{equation*}
    where $\mc{R}^\star(M_1, M_2)$ and
    $\mc{R}_C(M_1, M_2)$ are defined
    in~\eqref{Eqn:FeasibleRegion} and \eqref{Eqn:AchievableRatesG},
    respectively, and where $c_1$ and $c_2$ are finite positive
    constants independent of all the problem parameters.  
\end{theorem}

Theorem~\ref{Theorem:Bounds} shows that the region $\mc{R}_C(M_1,
M_2)$ is indeed feasible (since $\mc{R}_C(M_1, M_2) \subseteq
\mc{R}^\star(M_1, M_2)$). Moreover, the theorem shows that, up to a
constant additive and multiplicative gap, the scheme achieving
$\mc{R}_C(M_1, M_2)$ is optimal (since $\mc{R}^\star(M_1, M_2)
\subseteq  c_1 \cdot \mc{R}_C(M_1, M_2) - c_2$). From our analysis, we have the constants $c_1 = 1 / 60$ and $c_2 = 16$. However, numerical results suggest that the constants are in fact much smaller. 

The proof of Theorem~\ref{Theorem:Bounds} is presented in
Section~\ref{Sec:Proofs}. The proof actually
shows a slightly stronger result than stated in the theorem.  Recall
that the parameters $\alpha$ and $\beta$ control the weights of the
split between the two basic coded caching schemes mentioned above.  In
general, one would expect a tension between the rates
$R_1(\alpha,\beta)$ and $R_2(\alpha,\beta)$ over the first and second
hops of the network. In other words, the choice of $\alpha$ and $\beta$
minimizing the rate $R_1(\alpha,\beta)$ over the first hop will in
general \emph{not} minimize the rate $R_2(\alpha,\beta)$ over the second
hop. 

\begin{figure}[htbp]
    \centerline{\includegraphics{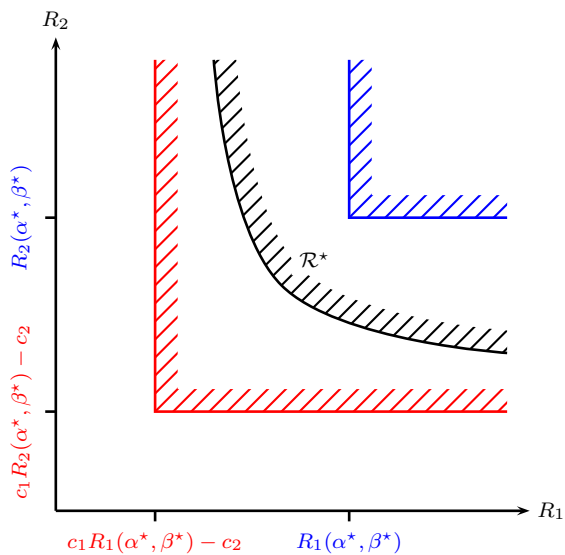}}   
    \caption{For fixed memory values $M_1$ and $M_2$, the figure
        qualitatively depicts the feasible rate region
        $\mc{R}^\star$
        and its bounds. As shown in
        the figure, the feasible rate region $\mc{R}^\star$ can be bounded
        by two rectangular regions with corner points 
        $\bigl( R_1(\alpha^\star, \beta^\star), R_2(\alpha^\star, \beta^\star) \bigr)$
        and 
        $\bigl( c_1R_1(\alpha^\star, \beta^\star)-c_2, c_1R_2(\alpha^\star,
        \beta^\star)-c_2 \bigr)$. Thus, up to the constant additive and
        multiplicative gap, there is no tension between the rates $R_1$
        over the first hop and the rate $R_2$ over the second hop of the
        optimal scheme for the hierarchical caching problem.}
    \label{Fig:Bounds}
\end{figure}

However, the proof of Theorem~\ref{Theorem:Bounds} shows that there
exists $\alpha^\star$ and $\beta^\star$ (depending on $N$, $M_1$, $M_2$,
$K_1$, and $K_2$) such that $R_1(\alpha^\star, \beta^\star)$ and
$R_2(\alpha^\star, \beta^\star)$ are \emph{simultaneously} approximately
minimized.  Thus, surprisingly, there is in fact \emph{no tension}
between the rates over the first hop and the second hop for the optimal
hierarchical caching scheme up to a constant additive and multiplicative
gap (see Fig.~\ref{Fig:Bounds}).  The next example shows that this is by
no means obvious, and, indeed, we conjecture that it is only true
approximately.

\begin{example}
    \label{eg:tension}
    Consider a setting with a single mirror $K_1 = 1$ and memory sizes
    $M_1 = M_2 = N/2$. Assume we use the proposed caching scheme with
    parameters $\alpha = 1/2$ and $\beta = 0$. As we will see later,
    this corresponds to placing one half of each file at the mirror and
    the other half at each of the caches. By~\eqref{Eqn:TotalRates}, we
    see that the rate tuple $(R_1, R_2) = (0, K_2/2)$ is achievable.
    Clearly, this minimizes the rate $R_1$ over the first hop. However,
    it is far from optimal for the second hop. 

Now, assume we use the proposed caching scheme with parameters
    $\alpha = \beta = 1/2$. By~\eqref{Eqn:TotalRates}, this achieves the
    rate tuple $(R_1, R_2) \approx (1/2,1)$. Observe that for an increase in
    rate of $1/2$ over the first link, we were able to decrease the rate of the second link 
    from $K/2$ to just one. 
    
    We conjecture that the rate tuple $(R_1, R_2) = (0,1)$ itself is not
    achievable.\footnote{This is because in order to achieve rate $0$
    over the first link, the mirror and each user together must store
    the entire content, which suggests that the cached contents of the
    mirror and each user do not overlap. However, to achieve rate $1$
    over the second link, the mirror needs to be able to exploit
    coded multicasting opportunities between the users, which
    suggests that the cached contents of the mirror and the users should
    overlap. This tension suggests that the rate tuple $(0,1)$ is not
    achievable.} If true, this implies that there is tension between the
    two rates but that this tension accounts for at most a constant
    additive and multiplicative gap. 
\end{example}

Before we provide the specific values of $\alpha^\star$ and $\beta^\star$, we
describe the two schemes controlled by these parameters in a bit more
detail. Both schemes make use of the coded caching scheme for networks
with a single layer of caches from \cite{CachingUM, DCachingUM} as
recalled in Section~\ref{Sec:SingleLayer}.

The first scheme uses a very natural decode-and-forward type approach.
It uses the single-layer scheme between the server and the $K_1$
mirrors. Each mirror decodes all messages for its children and
re-encodes them using the single-layer scheme between the mirror and its
$K_2$ attached users.  Thus, this first scheme creates and exploits
coded multicasting opportunities  between the server and the mirrors and
between each mirror and its users.  The second scheme simply ignores the
content of the mirrors and applies the single-layer scheme directly
between the server and the $K_1 K_2$ users.  Thus, this second scheme
creates and exploits coded multicasting opportunities between the server
and all the users.  With a choice of $(\alpha, \beta) = (1,1)$, all
weight is placed on the first scheme and the second scheme is not used.
With a choice of $(\alpha, \beta) = (0,0)$, all weight is placed on the
second scheme and the first scheme is not used. 

\begin{figure}[htbp]
    \centerline{\includegraphics{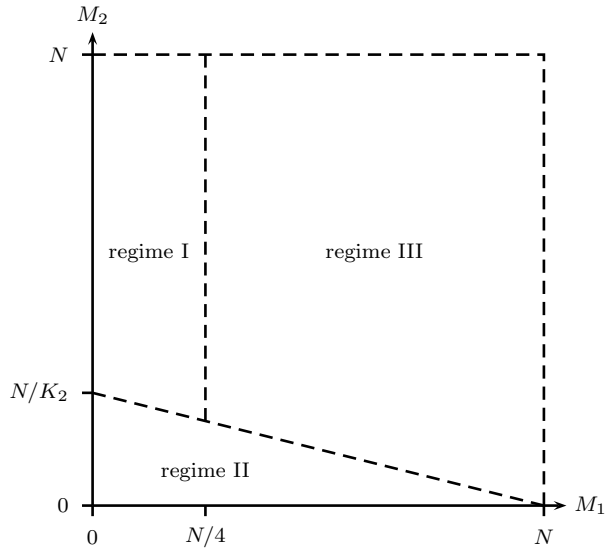}}
    \caption{Different regimes for $\alpha^\star$ and $\beta^\star$.} 
    \label{Fig:AlphaBeta0}
\end{figure}

With this in mind, let us return to the choice of $\alpha^\star$ and
$\beta^\star$. We consider the following three different regimes of $M_1$ and $M_2$, as
depicted in Fig.~\ref{Fig:AlphaBeta0}:
\begin{align}
\nonumber
   &\mbox{ I) Regime I: } M_1+M_2K_2 \geq N \mbox{ and } 0 \leq M_1 \leq N/4, \\
   \label{Eqn:Regimes}
    &\mbox{II) Regime II: } M_1+M_2K_2 < N, \\
    \nonumber
    &\mbox{III) Regime III: } M_1+M_2K_2 \geq N \mbox{ and } N/4 < M_1 \leq N.
\end{align}

We set 
\begin{equation}
    \label{Eqn:Optab0} 
    (\alpha^\star, \beta^\star) \defeq
    \begin{cases}
        \left(\displaystyle \frac{M_1}{N}, \frac{M_1}{N} \right) 
        & \text{in regime I}, \\[1em]
        \left( \displaystyle \frac{M_1}{M_1 + M_2K_2} , 0 \right)
        & \text{in regime II}, \\[1em]
        \left(\displaystyle \frac{M_1}{N}, \frac{1}{4} \right)
        & \text{in regime III}.
    \end{cases} 
\end{equation}
Substituting this choice into \eqref{Eqn:TotalRates}, the corresponding
achievable rates are
\begin{subequations}
    \label{Eqn:UpperBounds0}
    \begin{align}
        \label{Eqn:UpperBoundsR10}
        R_1(\alpha^\star , \beta^\star) 
        & \approx
        \begin{cases}
            \min \left\{ K_1K_2, \displaystyle \frac{N}{M_2} \right\} 
            & \text{in regime I},\\[1em] 
            \displaystyle \min \left\{ K_1K_2, \frac{M_1}{M_1 + M_2K_2} \cdot \frac{(N - M_1)K_2}{M_1 + M_2K_2} + \frac{M_2K_2}{M_1 + M_2K_2} \cdot \frac{NK_2 - M_1}{M_1 + M_2K_2} \right\} & \text{in regime II},\\[.85em]
            \displaystyle \frac{(N - M_1)^2}{NM_2} 
            & \text{in regime III},
        \end{cases} \\
        \shortintertext{and} \nonumber\\
        \label{Eqn:UpperBoundsR20}
        R_2(\alpha^\star , \beta^\star) 
        & \approx \min \left\{ K_2, \frac{N}{M_2} \right\},
    \end{align}
\end{subequations}
where the approximation is up to a constant additive and multiplicative
gap as before.

From~\eqref{Eqn:Optab0}, we see that in every regime we need to share
between the two basic schemes. In particular, using the natural
decode-and-forward type approach (i.e., scheme one) alone can be highly
suboptimal as the next two examples show.

\begin{example}
    \label{eg:df1}
    Let $M_1=0$ and $M_2=N$ so that the mirrors have zero memory and the
    user caches are able to store the entire database of files. This
    setting falls into regime I. We focus on the rate over the first
    link from the server to the mirrors. We know that in this example
    the optimal rate $R_1$ is $0$. By~\eqref{Eqn:UpperBoundsR10}, the
    rate $R_1(\alpha^\star, \beta^\star)$ is approximately equal to $1$ (a
    constant). However, the rate achieved by using only the
    first (decode-and-forward) scheme is equal to $R_1(1, 1) = K_1 K_2$,
    which could be much larger.
\end{example}

\begin{example}
    \label{eg:df2}
    Let $M_1 = N-N^{2/3}$, $M_2 = N^{1/4}$, $K_1 = 1$, and $K_2 =
    N^{5/6}$. This setting falls into regime III.
    By~\eqref{Eqn:UpperBoundsR10}, the rate $R_1(\alpha^\star, \beta^\star)$ is
    approximately equal to $N^{1/12}$. On the other hand, the rate
    achieved by using only the first (decode-and-forward) scheme is
    approximately equal to $N^{1/2}$, which could again be much larger.
\end{example}

\section{Caching Schemes} 
\label{Sec:AchievableSchemes}

In this section, we introduce a class of caching schemes for the
hierarchical caching problem.  We begin in Sections~\ref{Sec:SchemeA}
and~\ref{Sec:SchemeB} by using the BasePlacement and BaseDelivery
procedures defined in Section~\ref{Sec:SingleLayer} for networks with a
single layer of caches to construct two basic caching schemes for
networks with with two layers of caches.  We will see in
Section~\ref{Sec:SchemeG} how to combine these two schemes to yield a
near-optimal scheme for the hierarchical caching problem.

\subsection{Caching Scheme A}
\label{Sec:SchemeA}

\begin{figure}[htbp]
    \centerline{\includegraphics{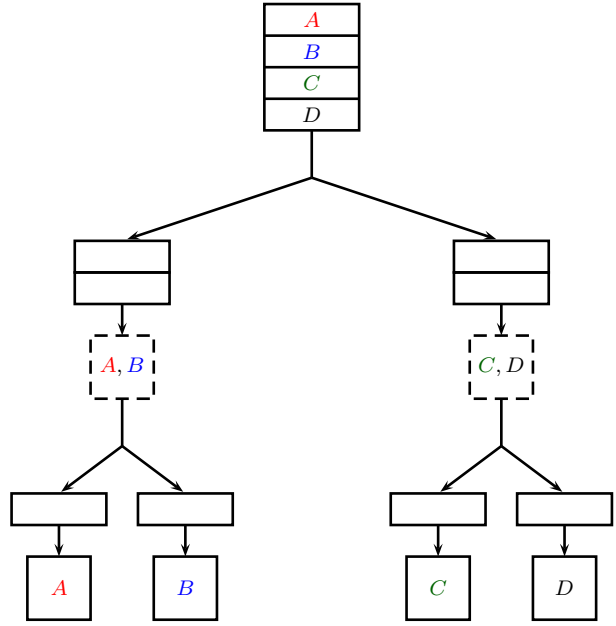}} 
    \caption{Caching scheme A for a system with $K_1 = 2$ mirrors and
        $K_2 = 2$ users per mirror. Scheme A uses a decode-and-forward
        type approach to apply the single-layer coded caching scheme
        recalled in Section~\ref{Sec:SingleLayer} to a network with two
        layers. We independently cache content in each of the layers
        during the placement phase. In the delivery phase, the mirrors
        decode all the files requested by their users and re-encode them for 
        their children. For example, in the figure mirror one decodes
        files $A$, $B$ and re-encodes them for the two attached users.}
    \label{Fig:SystemA}
\end{figure}

Informally, this scheme places content in the mirrors so that using the
server transmission and their own content, each mirror can recover all
the files requested by their attached users. In turn, each mirror then
acts as a server for these files. Content is stored in the attached user
caches so that from the mirror transmission and its own cache content,
each user can recover its requested file. See Fig.~\ref{Fig:SystemA} for
an illustration of the scheme. 

More formally, in the placement phase, we use the BasePlacement$(N, K_1,
M_1)$ procedure recalled in Section~\ref{Sec:SingleLayer} to store
portions of the files $1, 2, \ldots, N$ across the $K_1$
mirrors. Also, for each mirror $i$, we use the BasePlacement$(N, K_2,
M_2)$ procedure to independently store portions of the files $1, 2,
\dots, N$ across caches $(i, 1)$,  $(i, 2)$, \ldots, $(i, K_2)$
corresponding to the users with access to mirror $i$. In other words,
each mirror independently stores a random $M_1F/N$-bit subset of every
file, and each user cache independently stores a random $M_2F/N$-bit
subset of every file.

During the delivery phase, the server uses the BaseDelivery$(N, K_1,
M_1)$ procedure to the mirrors in order to enable them to recover the
$K_2$ files $d_{i, 1}, d_{i, 2}, \ldots, d_{i, K_2}$. In
other words, each mirror decodes all files requested by its attached
users.  Next, each mirror $i$ uses the BaseDelivery$(N, K_2, M_2)$
procedure to re-encode these files for its $K_1$ users. This enables
each user $(i,j)$ to recover its requested file $d_{i,j}$. Thus,
scheme A exploits coded multicasting opportunities between the server
and the mirrors and between the mirrors and their users.

The rates for caching scheme A are as follows.
By~\eqref{Eqn:BasicRateK2}, the rate over the link from the server to
the mirror is
\begin{subequations}
    \label{Eqn:AchievabilityA}
    \begin{align}
        \label{Eqn:AchievabilityA1}
        R_1^A & \defeq K_2 \cdot \rf\!\left( \frac{M_1}{N}, K_1\right).  \\
        \intertext{By~\eqref{Eqn:BasicRate}, the rate over the link from the mirrors to
        their users is}
        \label{Eqn:AchievabilityA2}
        R_2^A & \defeq \rf\!\left(\frac{M_2}{N}, K_2\right).
    \end{align}
\end{subequations}

\begin{example}
    \label{Ex:AchievableA}
    Consider the setup in Fig.~\ref{Fig:SystemA} with $N = 4$ files,
    $K_1 = 2$ mirrors, and $K_2 = 2$ users per mirror. The mirror and
    user cache memory sizes are $M_1 = 2$ and $M_2 = 1$, respectively.
    For ease of notation, denote the files by $A$, $B$, $C$ and
    $D$. Using scheme A, each mirror independently stores a
    random $F/2$-bit subset of every file, and each user cache
    independently stores a random $F/4$-bit subset of every file.

    In the delivery phase, assume the four users request files $A$, $B$,
    $C$, and $D$, respectively. The server uses the BaseDelivery
    procedure to enable the first mirror to recover files $A$ and $B$ and to
    enable the second mirror to recover files $C$ and $D$. This uses a rate of
    \begin{equation*}
        R_1^A = 2\cdot r(1/2, 2).
    \end{equation*}
    Mirror one then uses the BaseDelivery procedure to re-encode the files $A$
    and $B$ for its to attached users. Similarly, mirror two uses the
    BaseDelivery procedure to re-encode the files $C$ and $D$ for its attached
    users. This uses a rate of
    \begin{equation*}
        R_2^A = r(1/4, 2).
    \end{equation*}
\end{example}

\subsection{Caching Scheme B} 
\label{Sec:SchemeB}

\begin{figure}[htbp]
    \centerline{\includegraphics{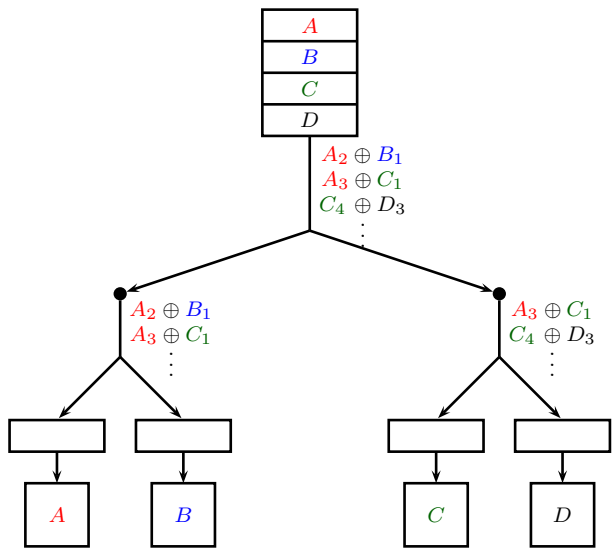}}   
    \caption{Caching scheme B for a system with $K_1 = 2$ mirrors
        and $K_2 = 2$ users per mirror. Scheme B ignores the memory at the
        mirrors and uses the single-layer coded caching scheme recalled in
        Section~\ref{Sec:SingleLayer} directly between the server and the
        users. The mirrors are only used to forward the relevant messages
        transmitted by the server to their users.}
    \label{Fig:SystemB}
\end{figure}

Informally, this scheme places content across the $K_1K_2$ user caches
so that using the server transmissions and its own cache content, each
user can recover its requested file. The storage capabilities of the
mirrors in the network are completely ignored and the mirrors are only
used to forward relevant parts of the server transmissions to the
corresponding users. See Fig.~\ref{Fig:SystemB} for an illustration.

More formally, in the placement phase, we use the BasePlacement$(N,
K_1K_2, M_2)$ procedure to store portions of the files $1, 2, \dots,
N$ across the $K_1K_2$ user caches and leave all the mirrors empty.
In other words, each user cache independently stores a random
$M_2F/N$-bit subset of every file.

During the delivery phase, the server uses the BaseDelivery$(N, K_1K_2,
M_2)$ procedure directly for the $K_1K_2$ users.  Recall from the
description in Section~\ref{Sec:SingleLayer} that the BaseDelivery
procedure transmits several sums of file parts. The transmission of
mirror $i$ consists of all those sums transmitted by the server that
involve at least one of the $K_2$ files $d_{i, 1}, d_{i, 2},
\ldots, d_{i, K_2}$, requested by its attached users $(i, 1)$, $(i,
2)$, \ldots, $(i, K_2)$.  From the information forwarded by the mirrors,
each user is able to recover its requested file. Thus, scheme B exploits
coded multicasting opportunities directly between the server and the
users across two layers.

The rates for caching scheme B are as follows. By~\eqref{Eqn:BasicRate},
the rate over the link from the server to the mirrors is 
\begin{subequations}
    \label{Eqn:AchievabilityB}
    \begin{align}
        \label{Eqn:AchievabilityB1}
        R_1^B & \defeq \rf\!\left(\frac{M_2}{N}, K_1K_2 \right). \\
        \intertext{Forwarding only the relevant server transmissions is
            shown in~\cite[Section~V.A]{DCachingUM} to result in a rate }
        \label{Eqn:AchievabilityB2}
        R_2^B & \defeq \rf\!\left(\frac{M_2}{N}, K_2 \right).
    \end{align}
\end{subequations}
between each mirror and its attached users.

\begin{example}
    \label{Ex:AchievableB}
    Consider the setup in Fig.~\ref{Fig:SystemB} with $N = 4$ files $K_1
    = 2$ mirrors, and $K_2 = 2$ users per mirror. The user cache memory
    size is $M_2 = 1$ (the mirror memory size $M_1$ is irrelevant here).
    For ease of notation, denote the files by $A$, $B$, $C$ and $D$.
    Furthermore, it will be convenient in the remainder of this example
    to label the users and caches as $1, 2, 3, 4$ as opposed to $(1,1),
    (1,2), (2,1), (2,2)$.  Using scheme B, each user cache independently
    stores a random $F/4$-bit subset of every file. 

    In the delivery phase, assume the four users request files $A$, $B$,
    $C$, and $D$, respectively. The server uses the BaseDelivery
    procedure to enable the users to recover their requested files as
    follows. Consider file $A$, and denote by $A_{\mc{S}}$ the bits of
    file $A$ stored exclusively at the user caches in $\mc{S}\subset
    \{1, 2, 3, 4\}$, and similarly for the other files $B, C, D$. The transmission from the server to the mirrors is
    then
    \begin{gather*}
        A_{2,3,4}\oplus B_{1,3,4}\oplus C_{1,2,4}\oplus D_{1,2,3} \\
        A_{2,3}\oplus B_{1,3}\oplus C_{1,2},\;
        A_{2,4}\oplus B_{1,4}\oplus D_{1,2},\;
        A_{3,4}\oplus C_{1,4}\oplus D_{1,3},\;
        B_{3,4}\oplus C_{2,4}\oplus D_{2,3} \\
        A_2\oplus B_1,\; 
        A_3\oplus C_1,\;
        A_4\oplus D_1,\; 
        B_3\oplus C_2,\;
        B_4\oplus D_2,\;
        C_4\oplus D_3 \\
        A_\emptyset,\;
        B_\emptyset,\;
        C_\emptyset,\;
        D_\emptyset.
    \end{gather*}
    For large enough file size $F$, this uses a normalized rate of
    \begin{equation*}
        R_1^B = r(1/4, 4).
    \end{equation*}

    Let us focus on mirror one. Since its attached users request files
    $A$ and $B$, it forwards every sum including parts of either of
    those files. Thus, mirror one transmits
    \begin{gather*}
        A_{2,3,4}\oplus B_{1,3,4}\oplus C_{1,2,4}\oplus D_{1,2,3} \\
        A_{2,3}\oplus B_{1,3}\oplus C_{1,2},\;
        A_{2,4}\oplus B_{1,4}\oplus D_{1,2},\;
        A_{3,4}\oplus C_{1,4}\oplus D_{1,3},\;
        B_{3,4}\oplus C_{2,4}\oplus D_{2,3} \\
        A_2\oplus B_1,\; 
        A_3\oplus C_1,\;
        A_4\oplus D_1,\; 
        B_3\oplus C_2,\;
        B_4\oplus D_2,\\
        A_\emptyset,\;
        B_\emptyset,\;
    \end{gather*}
    This uses a normalized rate of
    \begin{equation*}
        R_2^B = r(1/4, 2).
    \end{equation*}
\end{example}

\subsection{Generalized Caching Scheme}
\label{Sec:SchemeG}

\begin{figure}
    \centerline{\includegraphics{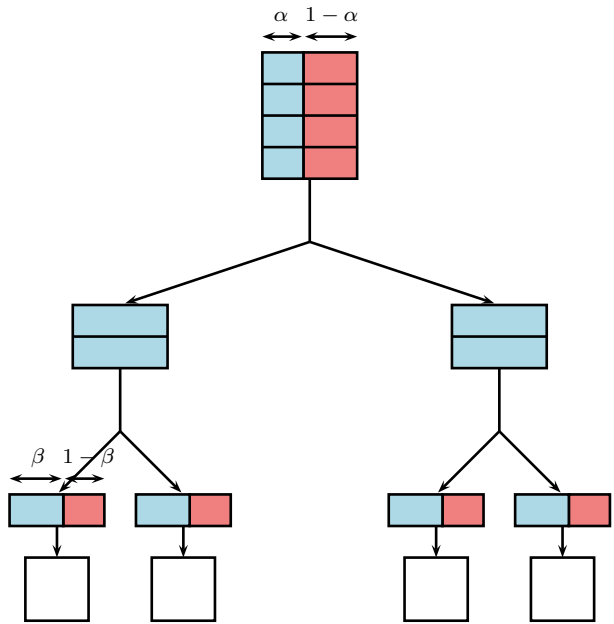}}   
    \caption{Generalized caching scheme for a system with $K_1=2$
        mirrors and $K_2 = 2$ users per mirror. For given $\alpha$ and
        $\beta$, the system is split into two disjoint subsystems.
        We use caching scheme A for delivering the first parts of the
        files over the first subsystem and use caching scheme B for
        delivering the second parts of the files over the second subsystem.}
    \label{Fig:Achievability}
\end{figure}

The generalized caching scheme divides the system into two subsystems, the first
one operated according to caching scheme A and the second one according
to caching scheme B. Fix parameters $\alpha, \beta  \in [0, 1]$. The
first subsystem includes the entire memory of each mirror and a $\beta$
fraction of each user cache memory. The second subsystem includes the
remaining $(1-\beta)$ fraction of each user cache memory. We split each
file into two parts of size $\alpha F$ and $(1-\alpha)F$ bits,
respectively. We use scheme A from Section~\ref{Sec:SchemeA} to store
and deliver the first parts of the files over the first subsystem.  Similarly, we use scheme B
from Section~\ref{Sec:SchemeB} for the second parts of the files over the second subsystem.  See
Fig.~\ref{Fig:Achievability} for an illustration.

Since our system is a composition of two disjoint subsystems, the net
rate over each transmission link is the sum of the corresponding rates in
the two subsystems. From~\eqref{Eqn:AchievabilityA}, the rates $R_1^1,
R_2^1$ required by scheme A over the first subsystem are
\begin{subequations}
    \label{Eqn:Subsystem1}
    \begin{align}
        R_1^1 & = \alpha K_2 \cdot \rf\!\left(\frac{M_1}{\alpha N}, K_1\right) , \\ 
        R_2^1 & =  \alpha \cdot \rf\!\left(\frac{\beta M_2}{\alpha N}, K_2 \right) .
    \end{align}
\end{subequations}
Similarly, from \eqref{Eqn:AchievabilityB}, the rates $R_1^2, R_2^2$
required by scheme B over the second subsystem are 
\begin{subequations}
    \label{Eqn:Subsystem2}
    \begin{align}
        R_1^2 & =  (1-\alpha) \cdot \rf\!\left(\frac{(1-\beta)M_2}{(1-\alpha)N}, K_1K_2 \right) , \\ 
        R_2^2 & = (1-\alpha) \cdot \rf\!\left(\frac{(1-\beta)M_2}{(1-\alpha)N}, K_2 \right) .
    \end{align}
\end{subequations}
The formal derivation for these rate expressions is provided in
Appendix~\ref{Sec:AppendixB}. 

Combining~\eqref{Eqn:Subsystem1} and~\eqref{Eqn:Subsystem2}, the net
rates $R_1 = R_1(\alpha,\beta)$ and $R_2 = R_2(\alpha,\beta)$ of the
generalized caching scheme are
\begin{subequations}
    \label{Eqn:NetRates}
    \begin{align}
        \label{Eqn:NetRates1}
        R_{1}(\alpha, \beta)
        & \defeq R_{1}^1 + R_{1}^{2} 
        = \alpha K_2 \cdot \rf\!\left(\frac{M_1}{\alpha N}, K_1\right) 
        + (1-\alpha) \cdot \rf\!\left(\frac{(1-\beta)M_2}{(1-\alpha)N}, K_1K_2 \right), \\
        \label{Eqn:NetRates2}
        R_2(\alpha,\beta)
        & \defeq
        R_{2}^1 + R_{2}^{2} 
        = \alpha \cdot \rf\!\left(\frac{\beta M_2}{\alpha N}, K_2 \right) 
        + (1-\alpha)\cdot  \rf\!\left(\frac{(1-\beta)M_2}{(1-\alpha)N}, K_2 \right).
    \end{align}
\end{subequations}
Note that this coincides with \eqref{Eqn:TotalRates}.

\subsection{Choice of $\alpha^\star$ and $\beta^\star$}
\label{Sec:Parameters}

The generalized caching scheme described in the last section is
parametrized by $\alpha$ and $\beta$. We now choose particular values
$\alpha^\star$ and $\beta^\star$ for these parameters.  Recall from \eqref{Eqn:Regimes} in 
Section~\ref{Sec:Results} the three regimes for the memory sizes $M_1$
and $M_2$:
\begin{enumerate}[I)]
    \item $M_1+M_2K_2 \geq N$ and $0 \leq M_1 \leq N/4$,
    \item $M_1+M_2K_2 < N$,
    \item $M_1+M_2K_2 \geq N$ and $N/4 < M_1 \leq N$.
\end{enumerate}
We set
\begin{equation}
    \label{Eqn:Optab1} 
    (\alpha^\star, \beta^\star) \defeq
    \begin{cases}
        \left(\displaystyle \frac{M_1}{N}, \frac{M_1}{N} \right) 
        & \text{in regime I}, \\[1em]
        \left( \displaystyle \frac{M_1}{M_1 + M_2K_2} , 0 \right)
        & \text{in regime II}, \\[1em]
        \left(\displaystyle \frac{M_1}{N}, \frac{1}{4} \right)
        & \text{in regime III}.
    \end{cases} 
\end{equation}
See also~\eqref{Eqn:Optab0}.

Our proof of Theorem~\ref{Theorem:Bounds} will demonstrate that for any
given $M_1, M_2$ and this choice of parameters $\alpha^\star,
\beta^\star$, the rates $R_1(\alpha^\star, \beta^\star),
R_2(\alpha^\star, \beta^\star)$ for the generalized caching scheme are
within a constant multiplicative and additive gap of the minimum
feasible rates. Before proceeding with the proof of this fact, we
provide intuition for the choice of these parameters as well as their
impact on the achievable scheme in each of these regimes.  

\begin{enumerate}[I)]
    \item We want to optimize the values of $\alpha, \beta$ with respect
        to both $R_1(\alpha, \beta), R_2(\alpha, \beta)$. Let us start
        with the rate $R_2(\alpha, \beta)$.
        From~\eqref{Eqn:AchievabilityA2}
        and~\eqref{Eqn:AchievabilityB2}, both caching schemes A and B
        achieve rate $R_2 = \rf(M_2/N, K_2)$ on the link from a mirror
        to its attached users. As we will see later, this rate is in
        fact approximately optimal for this link.  The generalized
        caching scheme combines caching schemes A and B, and it can be
        easily verified from \eqref{Eqn:NetRates2} that $\alpha = \beta$
        results in $R_2(\alpha, \beta) = \rf(M_2/N, K_2)$. Thus, $\alpha
        = \beta$ is near optimal with respect to $R_2(\alpha, \beta)$.
        To find the optimal common value, we analyze how the rate
        $R_1(\alpha, \alpha)$ varies with $\alpha$ and find that among
        all values in $[0,1]$, the choice $\alpha = M_1 / N$ results in
        the near-optimal rate for this regime. Thus, we choose
        $(\alpha^\star, \beta^\star) = (M_1 / N, M_1 / N)$ for this
        regime. 

        We now discuss the impact of this choice on the structure of the
        generalized caching scheme in this regime. Recall that caching
        scheme A is used to store and deliver the first parts of the
        files, each of size $\alpha^\star F$ bits. Since $\alpha^\star =
        M_1 / N$ and the mirror memory size is $M_1F$ bits, this implies
        that the \emph{entire} first parts of all the $N$ files can be
        stored in each mirror. Thus, in this regime, the server does not
        communicate with the mirrors regarding the first file parts and
        each mirror, in turn, acts as a sever for these files parts to
        its attached users.  Thus, the generalized caching scheme only
        exploits coded multicasting opportunities between each
        mirror and its attached users via caching scheme A and between
        the server and all the users via caching scheme B. 

    \item Observe that the user cache memory $M_2$ is small in this
        regime, in particular $M_2 < N / K_2$. It can be verified that
        the rate $R_2$ in this case has to be at least on the order of
        $K_2$. On the other hand, it is easy to see from
        \eqref{Eqn:NetRates2} that $R_2(\alpha, \beta) \le K_2$ for any
        choice of parameters $\alpha, \beta$. Thus, we only need to
        optimize $\alpha, \beta$ with respect to the rate $R_1(\alpha,
        \beta)$ over the second link. The optimizing values can be found
        as $(\alpha^\star, \beta^\star) = \left( M_1/ (M_1 + M_2K_2), 0
        \right)$. 

        Recall from Section~\ref{Sec:SchemeG} that caching scheme A is
        assigned a $\beta^\star$ fraction of each user cache memory.
        Since $\beta^\star = 0$ for this regime, no user cache memory is
        assigned for scheme A. Thus, in this regime, the generalized caching scheme only
        exploits coded multicasting opportunities between the server and
        its attached mirrors via caching scheme A and between the server
        and all the users via caching scheme B. 

    \item We would like to again choose $\alpha = \beta = M_1/N$ in this
        regime as in regime I.  However, since the rate $R_1(\alpha,
        \beta)$ over the first link increases with $\beta$, and since
        $M_1/N$ is large (on the order of $1$) in this regime, this
        choice would lead to an unacceptably large value of $R_1$.
        Thresholding $\beta$ at $1/4$ (or any other constant for that
        matter) in this regime enables us to simultaneously achieve
        the dual purpose of limiting its impact on the rate $R_1$,
        while still managing to reduce the rate $R_2$ sufficiently. Thus,
        for this regime we choose $(\alpha^\star, \beta^\star) = (M_1 /
        N, 1/4)$.   

        As was the case in regime I, since $\alpha^\star = M_1 / N$,
        each mirror is able to store the entire first parts of the $N$
        files and thus, the server does not communicate with the mirrors
        under caching scheme A. Thus, in this regime, the generalized
        caching scheme only exploits coded multicasting opportunities
        between each mirror and its attached users via caching scheme A
        and between the server and all the users via caching scheme B.  
\end{enumerate}

\subsection{Achievable rates $R_1(\alpha^\star, \beta^\star), R_2(\alpha^\star, \beta^\star)$}
\label{Sec:Rates}

We next calculate the achievable rates $R_1(\alpha^\star, \beta^\star)$
and $R_2(\alpha^\star, \beta^\star)$ of the generalized caching scheme
described in Section~\ref{Sec:SchemeG} with the choice of parameters
$\alpha^\star$ and $\beta^\star$ as given in
Section~\ref{Sec:Parameters}.

The achievable rates $R_1(\alpha, \beta),
R_2(\alpha, \beta)$ for the generalized caching scheme are given in
terms of the function $\rf(\cdot, \cdot)$, defined in
\eqref{Eqn:BasicRate}. It is easy to see that  
\begin{equation} 
    \label{Eqn:Rstar}
    \rf\!\left(\frac{M}{N}, K\right) \le
    \begin{cases}
        \min\left\{ K, \displaystyle \frac{N}{M} - 1\right\} & \text{for $M/N \le 1$,} \\
        0 & \text{otherwise.} 
    \end{cases}
\end{equation}

As defined in \eqref{Eqn:Optab1}, our choice of parameters
$(\alpha^\star, \beta^\star)$ takes different values for the three different
regimes of $M_1, M_2$. We evaluate the achievable rates
for each of these regimes. 

\medskip

I) $M_1+M_2K_2 \geq N$ and $0 \leq M_1 \leq N/4$. Recall
from~\eqref{Eqn:Optab1} that $(\alpha^\star , \beta^\star)  =
\left(M_1 / N , M_1 / N\right)$ in regime I.
From \eqref{Eqn:NetRates} and \eqref{Eqn:Rstar}, the achievable rates
$R_1(\alpha^\star , \beta^\star)$ and $R_2(\alpha^\star , \beta^\star)$
are upper bounded as
\begin{subequations}
    \label{Eqn:RA}
    \begin{align}
        \nonumber
        R_1(\alpha^\star, \beta^\star) 
        & =  \frac{M_1K_2}{N} \cdot \rf(1, K_1) + \left(1 - \frac{M_1}{N}\right) 
        \cdot \rf\!\left(\frac{M_2}{N}, K_1K_2 \right) \\
        \nonumber
        & \le 0 + \min\left\{ K_1K_2, \frac{N}{M_2} \right\} \\
        & = \min\left\{ K_1K_2, \frac{N}{M_2} \right\} , 
        \label{Eqn:R1A}
    \end{align}
    and
    \begin{align}
        \nonumber
        R_2(\alpha^\star, \beta^\star) 
        & = \frac{M_1}{N} \cdot \rf\!\left(\frac{M_2}{N} , K_2 \right) 
        + \left( 1 - \frac{M_1}{N} \right) \cdot \rf\!\left(\frac{M_2}{N} , K_2 \right) \\
        \nonumber
        & = \rf\!\left(\frac{M_2}{N} , K_2 \right) \\
        & \le \min \left\{ K_2, \displaystyle \frac{N}{M_2} \right\} . 
        \label{Eqn:R2A}
    \end{align}
\end{subequations}

\medskip

II) $M_1+M_2K_2 < N$.  Recall from \eqref{Eqn:Optab1} that
$(\alpha^\star , \beta^\star) = \left( M_1 / (M_1 + M_2K_2),
0\right)$ in regime II.  
From \eqref{Eqn:NetRates} and \eqref{Eqn:Rstar}, the achievable rate
$R_1(\alpha^\star , \beta^\star)$ is upper bounded as
\begin{subequations}
    \label{Eqn:RC}
    \begin{align}
        \nonumber
        R_1&(\alpha^\star, \beta^\star)  \\
        \nonumber
        & = \frac{M_1K_2}{M_1 + M_2K_2}\cdot \rf\!\left(\frac{M_1+ M_2K_2}{N}, K_1\right) 
        + \frac{M_2K_2}{M_1 + M_2K_2}\cdot \rf\!\left(\frac{M_1+M_2K_2}{NK_2}, K_1K_2 \right)\\
        \nonumber
        & \le \frac{M_1K_2}{M_1 + M_2K_2} 
        \cdot \min\left\{ K_1, \frac{N}{M_1 + M_2K_2} - 1\right\} 
        + \frac{M_2K_2}{M_1 + M_2K_2} 
        \cdot \min\left\{ K_1K_2, \frac{NK_2}{M_1 + M_2K_2} - 1\right\} \\
        \nonumber
        & \le \frac{M_1}{M_1 + M_2K_2} 
        \cdot \min\left\{K_1K_2, \frac{(N - M_1)K_2}{M_1 + M_2K_2} \right\} 
        + \frac{M_2K_2}{M_1 + M_2K_2} \cdot \min\left\{K_1K_2, \frac{NK_2 - M_1}{M_1 + M_2K_2}\right\} \\
        &\le \min \left\{ K_1K_2, \frac{M_1}{M_1 + M_2K_2} \cdot \frac{(N - M_1)K_2}{M_1 + M_2K_2} + \frac{M_2K_2}{M_1 + M_2K_2} \cdot \frac{NK_2 - M_1}{M_1 + M_2K_2} \right\}. 
        \label{Eqn:R1C}
    \end{align}
    For the first inequality we have used that $M_1 + M_2K_2 < N$ implies
    $M_1 < \alpha^\star N$ and $(1 - \beta^\star)M_2 = M_2 < (1 -
    \alpha^\star)N$ in the bound \eqref{Eqn:Rstar}.  On the other hand, from
    \eqref{Eqn:NetRates} and \eqref{Eqn:Rstar} the achievable rate
    $R_2(\alpha^\star , \beta^\star)$ is trivially upper bounded as
    \begin{equation}
        \label{Eqn:R2C}
        R_2(\alpha^\star, \beta^\star) \leq K_2 = \min\left\{K_2, \frac{N}{M_2} \right\}
    \end{equation}
\end{subequations}
where the last equality follows since $M_2K_2 < N$ in regime~II.  
\medskip

III) $M_1+M_2K_2 \geq N$ and $N/4 < M_1 \leq N$. Recall
from~\eqref{Eqn:Optab1} that $(\alpha^\star , \beta^\star)  = \left(
M_1 / N , 1 / 4\right)$ in regime III.  From
\eqref{Eqn:NetRates} and \eqref{Eqn:Rstar}, the achievable rates
$R_1(\alpha^\star , \beta^\star)$ and $R_2(\alpha^\star , \beta^\star)$
are upper bounded as 
\begin{subequations}
    \label{Eqn:RB}
    \begin{align}
        \nonumber
        R_1(\alpha^\star, \beta^\star) 
        & =  \frac{M_1K_2}{N}\cdot \rf\!\left(1, K_1\right) 
        + \left( 1 - \frac{M_1}{N} \right)
        \cdot \rf\!\left(\frac{( 1 - 1/4)M_2}{( 1 - M_1/N)N}, K_1K_2 \right) \\
        \nonumber
        &\le 0 + \left( 1 - \frac{M_1}{N} \right)
        \cdot  \min\left\{ K_1K_2, \frac{4(N- M_1)}{3M_2} - 1 \right\} \\
        &\le \frac{4(N - M_1)^2}{3NM_2} 
        \label{Eqn:R1B} 
    \end{align}
    and
    \begin{align}
        \nonumber
        R_2(\alpha^\star, \beta^\star) 
        & = \frac{M_1}{N} \cdot \rf\!\left(\frac{M_2}{4M_1}, K_2\right) 
        + \left(1- \frac{M_1}{N} \right) 
        \cdot \rf\!\left(\frac{3M_2}{4(N-M_1)}, K_2 \right)\\
        \nonumber
        &\le \frac{M_1}{N} \cdot \min\left\{ K_2, \frac{4M_1}{M_2} \right\} + \left(1- \frac{M_1}{N} \right)\cdot  \min \left\{ K_2, \frac{4(N - M_1)}{3M_2} \right\} \\
        \nonumber
        & \le  \frac{M_1}{N} \cdot \min\left\{ K_2, \frac{4N}{M_2} \right\} 
        + \left(1- \frac{M_1}{N} \right)\cdot \min \left\{ K_2, \frac{4N}{M_2} \right\} \\
        & \leq 4 \min \left\{ K_2, \frac{N}{M_2} \right\}.
        \label{Eqn:R2B} 
    \end{align}
\end{subequations}

\medskip

Combining \eqref{Eqn:R1A}, \eqref{Eqn:R1C}, and \eqref{Eqn:R1B}, we
obtain the following upper bound on the achievable rate
$R_1(\alpha^\star , \beta^\star)$:
\begin{subequations}
    \label{Eqn:UpperBounds}
    \begin{equation}
        R_1(\alpha^\star , \beta^\star) \le
        \begin{cases}
            \min \left\{ K_1K_2, \displaystyle \frac{N}{M_2} \right\} &
            \text{in regime I}, \\[.85em]
            \displaystyle \min \left\{ K_1K_2, \frac{M_1}{M_1 + M_2K_2} \cdot \frac{(N - M_1)K_2}{M_1 + M_2K_2} + \frac{M_2K_2}{M_1 + M_2K_2} \cdot \frac{NK_2 - M_1}{M_1 + M_2K_2} \right\} & \text{in regime II},\\[.85em]
            \displaystyle \frac{4(N - M_1)^2}{3NM_2} & \text{in regime III}. 
        \end{cases}
        \label{Eqn:UpperBoundsR1}
    \end{equation}
    Similarly, combining \eqref{Eqn:R2A}, \eqref{Eqn:R2C}, and
    \eqref{Eqn:R2B}, we obtain the following upper bound on the achievable rate
    $R_2(\alpha^\star , \beta^\star)$:
    \begin{align}
        R_2(\alpha^\star , \beta^\star) &\le 4 \min \left\{ K_2, \frac{N}{M_2} \right\} .
        \label{Eqn:UpperBoundsR2}
    \end{align}
\end{subequations}
These upper bounds will be used in the next sections to prove that the
achievable rates for our generalized caching scheme are within a
constant multiplicative and additive gap of the corresponding lower
bounds.

\section{Proof of Theorem~\ref{Theorem:Bounds}}
\label{Sec:Proofs}

\subsection{Proof of $\mc{R}_C(M_1, M_2) \subseteq \mc{R}^\star(M_1, M_2)$}

Recall the definitions of the feasible rate region $\mc{R}^\star(M_1,
M_2)$ in~\eqref{Eqn:FeasibleRegion} and of the region $\mc{R}_C(M_1,
M_2)$ in~\eqref{Eqn:AchievableRatesG}, respectively. The result then
follows immediately from \eqref{Eqn:NetRates} in
Section~\ref{Sec:SchemeG}, which shows that any rate pair in
$\mc{R}_C(M_1, M_2)$ is achievable using the generalized caching scheme.
\hfill\IEEEQED

\subsection{Proof of $\mc{R}^\star(M_1, M_2) \subseteq  c_1 \cdot \mc{R}_C(M_1, M_2) - c_2$}
\label{Sec:Proofs_subset}

The proof consists of two steps. We first prove lower bounds
$R_1^{\textup{lb}}\left(M_1, M_2\right),  R_2^{\textup{lb}}\left(M_1, M_2\right)$ on the
feasible rates, i.e., for any $M_1, M_2$, and $(R_1, R_2) \in
\mc{R}^\star(M_1, M_2)$, we have
\begin{align}
    \nonumber
    R_1 &\ge R_1^{\textup{lb}}\left(M_1, M_2\right), \\
    R_2 &\ge R_2^{\textup{lb}}\left(M_1, M_2\right) .
    \label{Eqn:Proof1}
\end{align}
We compute these lower bounds $R_1^{\textup{lb}}\left(M_1, M_2\right),
R_2^{\textup{lb}}\left(M_1, M_2\right)$ in Appendix~\ref{Sec:LowerBounds}. 

Next, we show that for any $M_1, M_2$, the gap between the achievable
rates $R_1(\alpha^\star, \beta^\star), R_2(\alpha^\star, \beta^\star)$
and the lower bounds $R_1^{\textup{lb}}\left(M_1, M_2\right),  R_2^{\textup{lb}}\left(M_1,
M_2\right)$ is bounded, i.e., 
\begin{align}
    \nonumber
    R_1^{\textup{lb}}\left(M_1, M_2\right) &\ge c_1 R_1(\alpha^\star, \beta^\star) - c_2 ,\\
    R_2^{\textup{lb}}\left(M_1, M_2\right) &\ge c_1 R_2(\alpha^\star, \beta^\star) - c_2 ,
    \label{Eqn:Proof2}
\end{align}
where $c_1, c_2$ are finite positive constants independent of all the
problem parameters. The proof of the above inequalities, bounding the
gap between the achievable rates and the lower bounds, involves separate
analysis for several different regimes of $M_1, M_2$, and is deferred to
Appendices~\ref{Sec:GapR1} and~\ref{Sec:GapR2}. 

Combining \eqref{Eqn:Proof1}, \eqref{Eqn:Proof2}, for any $M_1, M_2$,
and $(R_1, R_2) \in \mc{R}^\star(M_1, M_2)$, we have 
\begin{align*}
    R_1 &\ge c_1 R_1(\alpha^\star, \beta^\star) - c_2 ,\\
    R_2 &\ge c_1 R_2(\alpha^\star, \beta^\star) - c_2 . 
\end{align*}
Since $\mc{R}_C(M_1, M_2)$ is precisely the set of tuples of the form
$(R_1(\alpha, \beta), R_2(\alpha, \beta))$ for some $\alpha,
\beta\in[0,1]$, this shows that
$\mc{R}^\star(M_1, M_2) \subseteq  c_1 \cdot \mc{R}_C(M_1, M_2) - c_2$,
completing the proof. \hfill\IEEEQED

As mentioned in Section~\ref{Sec:Results}, the proof above shows a
stronger result than claimed in the theorem statement. In particular, it
shows that for any $M_1$ and $M_2$ there exists parameters
$\alpha^\star$ and $\beta^\star$ such that both $R_1(\alpha^\star,
\beta^\star)$ and $R_2(\alpha^\star, \beta^\star)$ are
\emph{simultaneously} approximately close to their minimum value.  In
other words, up to a constant additive and multiplicative gap, there is
no tension between the rates over the first and second hops of the
network for the optimal caching scheme. 

\appendices

\section{Rates for the Generalized Caching Scheme}
\label{Sec:AppendixB}

This appendix derives the rate expressions \eqref{Eqn:Subsystem1} and
\eqref{Eqn:Subsystem2} in Section~\ref{Sec:SchemeG} for the two
subsystems using the generalized caching scheme. 

Recall that the first subsystem is concerned with caching and delivering
the first $\alpha$ fraction of each file. It includes the entire memory
of each mirror and the first $\beta$ fraction of each user cache. Let 
\begin{align*}
    F^1 & \defeq \alpha F, \\
    M_1^1 & \defeq  \frac{M_1 F}{F^1} = \frac{M_1}{\alpha} , \\  
    M_2^1 & \defeq  \frac{\beta M_2 F}{F^1} = \frac{\beta M_2}{\alpha} 
\end{align*}
denote the equivalent file size, as well as mirror memory and user cache memory, normalized by the equivalent file size, 
for this subsystem. From \eqref{Eqn:AchievabilityA}, the rates $R_1^1,
R_2^1$ (normalized by the file size $F$) required by caching scheme A on
this subsystem are given by 
\begin{align*}
    R_1^1 & = \alpha K_2 \rf\!\left(\frac{M_1^1}{N} , K_1\right) 
    = \alpha K_2 \rf\!\left(\frac{M_1}{\alpha N}, K_1\right) , \\ 
    R_2^1 & = \alpha \rf\!\left(\frac{M_2^1}{N}, K_2\right) 
    = \alpha \rf\!\left(\frac{\beta M_2}{\alpha N}, K_2 \right) .
\end{align*}

The second subsystem is concerned with caching and delivering the second
$1-\alpha$ fraction of each file. It only uses the memory in the second
$1-\beta$ fraction of each user cache. Let
\begin{align*}
    F^2 & \defeq (1-\alpha) F, \\
    M_2^2 & \defeq  \frac{(1-\beta)M_2 F}{F^2} = \frac{(1-\beta)M_2}{(1 - \alpha)} 
\end{align*}
denote the equivalent file size and user cache memory, normalized by the equivalent file size,  for this
subsystem. From \eqref{Eqn:AchievabilityB}, the rates $R_1^2, R_2^2$
(again normalized by the file size $F$) required by caching scheme B on
this subsystem are given by 
\begin{align*}
    R_1^2 & = (1-\alpha) \rf\!\left(\frac{M_2^2}{N}, K_1K_2\right) 
    = (1-\alpha) \rf\!\left(\frac{(1-\beta)M_2}{(1-\alpha)N}, K_1K_2 \right) , \\ 
    R_2^2 & = (1-\alpha) \rf\!\left( \frac{M_2^2}{N}, K_2\right) 
    = (1-\alpha) \rf\!\left(\frac{(1-\beta)M_2}{(1-\alpha)N}, K_2 \right) .
\end{align*}

\section{Lower bounds}
\label{Sec:LowerBounds}

Given any $M_1, M_2$, we want to establish lower bounds on the rates
$R_{1}, R_{2}$ for the tuple $\left(M_1, M_2, R_1, R_2\right)$ to be
achievable. Our lower bounds are similar to the one proposed in
\cite{CachingUM} for single-layer caching networks. 

Assume the tuple $\left(M_1, M_2, R_1, R_2\right)$ is feasible and
consider the shared communication link between the server and the
mirrors. Fix $s_1 \in \{1,2,\ldots, K_1\}$ and $s_{2} \in \{1,2,\ldots
K_2\}$. Consider the set of $s_1\cdot s_2$ users $(i,j)$ with
$i\in\{1,2,\dots,s_1\}$ and $j\in\{1,2,\dots,s_2\}$.  Consider a request
matrix $\bm{D}$ with user $(i, j)$ requesting $d_{i,j} = (i-1)s_2 + j$.
Since the tuple $\left(M_1, M_2, R_1, R_2\right)$ is feasible, each user
$(i, j)$ can recover its requested file from the transmission from the
server of rate $R_1$ along with the contents of mirror $i$ of size $M_1$
and cache $(i, j)$ of size $M_2$.

Now, consider a different request matrix $\bm{D}$ in which user 
$(i, j)$ requests $d_{i,j} = s_1s_2+(i-1)s_2 + j$. Again from the server
transmission of rate $R_1$ and the two cache memories of sizes $M_1$ and
$M_2$ each user $(i,j)$ can recover its requested file. Note that, while
the transmission of the server can depend on the request matrix, the
contents of the caches do not.

Repeat the same argument for a total of $\left\lfloor N / (s_1s_2)
\right\rfloor$ request matrices.  Then we have the following cut-set
bound \cite{Cover91}:
\begin{equation}
    \label{Eqn:LowerBound1}
    \left\lfloor \frac{N}{s_1s_2} \right\rfloor R_1 + s_1 M_1 + s_1s_2
    M_2 \ge \left\lfloor \frac{N}{s_1s_2} \right\rfloor s_1s_2.
\end{equation}
On the left-hand side of~\eqref{Eqn:LowerBound1}, the first term
corresponds to the $\left\lfloor N / (s_1s_2) \right\rfloor$
transmissions from the server, one for each request matrix, of rate
$R_1$ each; the second term corresponds to the $s_1$ mirror memories;
and the third term corresponds to the $s_1s_2$ user memories.  The
right-hand side of~\eqref{Eqn:LowerBound1} corresponds to the $s_1s_2$
different files that are reconstructed by the users for each of the
$\left\lfloor N / (s_1s_2) \right\rfloor$ request matrices.
\eqref{Eqn:LowerBound1} can be rewritten as
\begin{align}
\nonumber
       R_1 &\ge s_1s_2 - \frac{s_1 M_1+s_1s_2M_2}{ \left\lfloor N/(s_1s_2) \right\rfloor } \\
\nonumber
&\ge s_1s_2 - \frac{s_1 M_1}{N / (s_1s_2) - 1 } - \frac{s_1s_2M_2}{  N / (s_1s_2) - 1 } \\
&= s_1s_2 \left( 1 - \frac{ s_1M_1 + s_1s_2M_2}{N - s_1s_2} \right). 
  \label{Eqn:R1LB1}   
\end{align}

We can modify the above argument slightly to get an alternate lower
bound on the rate $R_1$. Instead of $\left\lfloor N / (s_1s_2)
\right\rfloor$ transmissions, we will use $\left\lceil N / (s_1s_2)
\right\rceil$ transmissions in \eqref{Eqn:LowerBound1} to get
\begin{equation*}
    \left\lceil \frac{N}{s_1s_2} \right\rceil R_1 + s_1 M_1 + s_1s_2 M_2 
    \ge N,
\end{equation*}
or, equivalently, 
\begin{align}
\nonumber
R_1 &\ge \frac{N - s_1M_1 - s_1s_2M_2}{ \left\lceil N/(s_1s_2) \right\rceil } \\
\nonumber
   &\ge \frac{N - s_1M_1 - s_1s_2M_2}{ N / (s_1s_2)  + 1 } \\
   &= \frac{s_1s_2\left(N - s_1M_1 - s_1s_2M_2\right)}{N + s_1s_2}. 
    \label{Eqn:R1LB2}
\end{align}

Since the inequalities \eqref{Eqn:R1LB1} and \eqref{Eqn:R1LB2} hold true
for any choice of $s_1 \in \{1,2,\ldots, K_1\}$ and
$s_{2} \in \{1,2,\ldots K_2\}$, we have the following lower bound on the
rate $R_1$ for the tuple $\left(M_1, M_2, R_1, R_2\right)$ to be
feasible:
\begin{align}
    \nonumber
    R_1 
    & \ge \max_{ \substack{s_1 \in \{1,2,\ldots, K_1\} \\ s_2 \in \{1,2\ldots, K_2\} } } 
    \max\left\{  s_1s_2 \left( 1 - \frac{ s_1M_1 + s_1s_2M_2}{N - s_1s_2} \right), 
    \ \ \frac{s_1s_2\left(N - s_1M_1 - s_1s_2M_2\right)}{N + s_1s_2} \right\}\\
    & \defeq R_{1}^{\textup{lb}}\left(M_1, M_2\right).
    \label{Eqn:NetLowerBoundR1}
\end{align}

\subsection{Rate $R_2$}

Assume the tuple $\left(M_1, M_2, R_1, R_2\right)$ is feasible and
consider the link between mirror one and its attached users. Let $t \in
\{1,2,\ldots K_2\}$. Consider the set of $t$ users $(1,j)$ with
$j\in\{1,2,\dots,t\}$.  Consider a request matrix $\bm{D}$ with user
$(1,j)$ requesting $d_{1,j} =  j$. Since the tuple $\left(M_1, M_2, R_1,
R_2\right)$ is feasible, each user $(1, j)$ can recover its requested
file  from the message transmitted by mirror one of rate $R_2$ and the
contents of its cache of size $M_2$. 

Now, consider a different request matrix $\bm{D}$ in which user 
$(1, j)$ requests $d_{i,j} = t + j$. Again from the mirror
transmission of rate $R_2$ and its cache of size $M_2$ 
each user $(1,j)$ can recover its requested file. Note that, while
the transmission of the mirror can depend on the request matrix, the
contents of the caches do not.

Repeat the same argument for a total of $\left\lceil N / t
\right\rceil$ request matrices. Then we have the following cut-set
bound \cite{Cover91}:
\begin{equation*}
    \left\lceil \frac{N}{t} \right\rceil R_2 +  t M_2 \ge N,  
\end{equation*}
or, equivalently, 
\begin{align*}
    R_2 & \ge  \frac{N - tM_2}{ \left\lceil N/t \right\rceil } .
\end{align*}

Since this inequality holds true for any choice of $t \in \{1,2,\ldots
K_2\}$, we have the following lower bound on the rate $R_2$ for the
tuple $\left(M_1, M_2, R_1, R_2\right)$ to be feasible:
\begin{align}
    \nonumber
    R_2 &\ge \max_{t \in \{1,2\ldots, K_2\} } \  \frac{N - tM_2}{ \left\lceil N/t \right\rceil }  \\
    \nonumber
&\ge \max_{t \in \{1,2\ldots, K_2\} }   \  \frac{N - tM_2}{\frac{N}{t} + 1 } \\
\nonumber
&= \max_{t \in \{1,2\ldots, K_2\} }   \ \frac{t(N - tM_2)}{ N + t} \\
    & \defeq R_{2}^{\textup{lb}}\left(M_1, M_2\right).
    \label{Eqn:LowerBoundR2}
\end{align}

\section{Gap between achievable rate $R_1(\alpha^\star, \beta^\star)$ and lower bound $R_{1}^{\textup{lb}}\left(M_1, M_2\right)$} 
\label{Sec:GapR1}

\Roman{subsection}
\Alph{subsubsection}
\renewcommand{\thesubsectiondis}{\textup{\Roman{subsection}}.}
\renewcommand{\thesubsubsectiondis}{\textup{\thesubsectiondis\Alph{subsubsection}})}
\renewcommand{\thesubsection}{\textup{\Roman{subsection}}}
\renewcommand{\thesubsubsection}{\textup{\thesubsection.\Alph{subsubsection}}}

We prove that the rate $R_1(\alpha^\star, \beta^\star)$ over the first hop, for the generalized caching scheme, as described in \eqref{Eqn:UpperBoundsR1} is within a constant additive and multiplicative gap of the minimum feasible rate $R_1$ for all values of $M_1, M_2$. Recall from \eqref{Eqn:Optab0} and Fig.~\ref{Fig:AlphaBeta0} that we use different parameters $(\alpha^\star, \beta^\star)$ for the generalized caching scheme in the three different regimes of $(M_1, M_2)$, regimes~I, II, and III. To prove the result, we will consider each of these regimes of $(M_1, M_2)$ in sequence, and bound the gap between the achievable rate $R_1(\alpha^\star, \beta^\star)$ and the corresponding lower bound $R_1^{\textup{lb}}(M_1, M_2)$, as derived in Appendix~\ref{Sec:LowerBounds}. Henceforth, we focus on the case where $K_1, K_2 \ge 4$. For $K_1 \leq 3$ ($K_2 \leq 3$), it is easy to see that the optimal rate is  within the constant factor $3$ of the rate of the network with $K_1 = 1$ ($K_2 = 1$). The optimum rate for $K_1=1$ ($K_2 = 1$) can be characterized easily following the results of \cite{CachingUM}.

We begin with regime~I.

\subsection*{\textup{Regime I:} $\displaystyle M_1 + M_2K_2 \ge N, \ 0 \le M_1 < \frac{N}{4}$}
\addtocounter{subsection}{1}
\label{Sec:BGRegimeI}
For this regime, recall from \eqref{Eqn:UpperBoundsR1} that the achievable rate $R_1(\alpha^\star, \beta^\star)$ is upper bounded as
\begin{equation}
    \label{Eqn:RateR1Reg1}
    R_1(\alpha^\star, \beta^\star) \le \min\left\{ K_1K_2, \frac{N}{M_2} \right\} . 
\end{equation}
On the other hand, recall the following lower bound on the rate $R_1$ from \eqref{Eqn:NetLowerBoundR1}: 
\begin{align}
    R_1^{\textup{lb}}(M_1, M_2) 
    &\ge  \max_{ \substack{s_1 \in \{1,2,\ldots, K_1\} \\ s_2 \in \{1,2\ldots, K_2\} } }   \ \frac{s_1s_2\left(N - s_1M_1 - s_1s_2M_2\right)}{N + s_1s_2} 
    \label{Eqn:LowerBoundR1App2}
\end{align}
For characterizing the gap between the achievable rate and the lower
bound, we further divide this regime into three subregimes as follows: 
\begin{alignat*}{2}
    &\hspace{-2in}\textup{\ref{Sec:RegimeICaseA})}& \ \displaystyle 0 \le M_1 < \frac{N}{2K_1}&, \ \ \frac{3N}{4K_2} \le M_2 < \frac{N}{4}, \\[1em]
    &\hspace{-2in}\textup{\ref{Sec:RegimeICaseB})}& \ \displaystyle \frac{N}{2K_1} \le M_1 < \frac{N}{4}&, \ \ \frac{3N}{4K_2} \le M_2 < \frac{N}{4}, \\[1em]
    &\hspace{-2in}\textup{\ref{Sec:RegimeICaseC})}& \ \displaystyle 0 \le M_1 < \frac{N}{4}&, \ \ \frac{N}{4} \le M_2 \le N. 
\end{alignat*}
The subregimes above only consider $M_2 \ge 3N / (4K_2)$ since for
regime~I, we have $M_1 + M_2K_2 \ge N$ and $M_1 < N / 4$, and thus $M_2
\ge (N - M_1) / K_2 \ge 3N / (4K_2)$. We now consider the three
subregimes one by one. 

\bigskip

\subsubsection{$\displaystyle 0 \le M_1 < \frac{N}{2K_1}, \ \ \frac{3N}{4K_2} \le M_2 < \frac{N}{4}$} 
\label{Sec:RegimeICaseA} 

Let 
\begin{align*}
    s_1 & = 1, \\
    s_2 & = \left\lfloor \frac{N}{2M_2} \right\rfloor 
\end{align*}
in the lower bound in \eqref{Eqn:LowerBoundR1App2}. Using $\lfloor x
\rfloor \ge x / 2$ for any $x \ge 1$, we can confirm that this is a
valid choice since 
\begin{equation} 
    \label{Eqn:Temp1} 
    1 \le \frac{N}{4M_2} 
    \le \left\lfloor \frac{N}{2M_2} \right\rfloor  
    \le \frac{N}{2M_2} 
    \le \frac{2K_2}{3}.  
\end{equation}
Then, by evaluating \eqref{Eqn:LowerBoundR1App2} we have 
\begin{align*} 
    R_1^{\textup{lb}}\left(M_1, M_2\right)  
    &\ge \frac{ \left\lfloor \frac{N}{2M_2} \right\rfloor \left(N - M_1 -
    \left\lfloor \frac{N}{2M_2} \right\rfloor  M_2 \right)}
    {N + \left\lfloor \frac{N}{2M_2} \right\rfloor} \\
    & \overset{(a)}{\ge} \frac{ \frac{N}{4M_2} \left( N - \frac{N}{2K_1} - \frac{N}{2M_2}  M_2 \right)}
    {N + \frac{N}{2M_2}} \\
    &\overset{(b)}{\ge} \frac{6N}{4 \cdot 7 \cdot M_2} \left( 1 - \frac{1}{2K_1} - \frac{1}{2} \right) \\ 
    &\overset{(c)}{\ge} \frac{3N}{14 M_2} \left( \frac{1}{2} - \frac{1}{8} \right) \\ 
    &\ge \frac{N}{13M_2} \\ &\ge \frac{1}{13} \min \left\{ K_1K_2, \frac{N}{M_2} \right\} 
\end{align*}
where $(a)$ follows since $\lfloor x \rfloor \ge x / 2$ for any $x \ge
1$; $(b)$ follows since $$ \frac{N}{2M_2} \le \frac{2K_2}{3} =
\frac{2K_1K_2}{3K_1} \le \frac{2N}{3K_1} \le \frac{N}{6} $$ 
using \eqref{Eqn:Temp1},  $N \ge K_1K_2$, and $K_1 \ge 4$; and $(c)$
follows since we have $K_1 \ge 4$. Combining with
\eqref{Eqn:RateR1Reg1}, we have 
\begin{equation} 
    \label{Eqn:BGRegimeICaseA} 
    R_1^{\textup{lb}}\left(M_1, M_2\right) 
    \ge \frac{1}{13} R_1(\alpha^\star, \beta^\star) .
\end{equation}

\bigskip

\subsubsection{$\displaystyle \frac{N}{2K_1} \le M_1 < \frac{N}{4}, \ \ \frac{3N}{4K_2} \le M_2 < \frac{N}{4}$} 
\label{Sec:RegimeICaseB}

Let 
\begin{align*}
    (s_1 , s_2)  & = 
    \begin{cases}
        \left( \left\lfloor\frac{N}{4M_1} \right\rfloor, \left\lfloor \frac{M_1}{M_2} \right\rfloor \right) &\mbox{ if } M_1 \ge M_2, \\[.75em]
        \left( \left\lfloor \frac{N}{4M_2}\right\rfloor, 1 \right)  &\mbox{ otherwise,}  
    \end{cases}
\end{align*}
in \eqref{Eqn:LowerBoundR1App2}. This is a valid choice since for $M_1 \ge M_2$, we have
\begin{align*}
    1 =  
    \left\lfloor\frac{N}{4 \cdot N / 4} \right\rfloor 
    \le  &\left\lfloor\frac{N}{4M_1} \right\rfloor 
    \le \frac{N}{4M_1} 
    \le \frac{K_1}{2} ,\\[.75em]
    1 \le 
    & \left\lfloor \frac{M_1}{M_2} \right\rfloor  
    \le \frac{M_1}{M_2} 
    \le \frac{N/ 4}{3 N / (4K_2)} 
    = \frac{K_2}{3},
\end{align*}
and for $M_1 < M_2$, we have 
\begin{align*}
    1 = 
    \left\lfloor\frac{N}{4 \cdot N / 4} \right\rfloor 
    \le  &\left\lfloor\frac{N}{4M_2} \right\rfloor 
    \le \left\lfloor\frac{N}{4M_1} \right\rfloor 
    \le \frac{N}{4M_1} 
    \le \frac{K_1}{2}.
\end{align*}
Note that $s_1 \le N / (4M_1)$ and $s_1s_2 \le N / (4M_2)$. Further,
since $\lfloor x \rfloor \ge x / 2$ for any $x \ge 1$, we have $s_1s_2
\ge N / (16M_2)$. Finally, substituting $s_1, s_2$ in
\eqref{Eqn:LowerBoundR1App2}, we obtain 
\begin{align*}
    R_{1}^{\textup{lb}}\left(M_1, M_2\right) &\ge \frac{\frac{N}{16M_2} \left(N - \frac{N}{4M_1} \cdot M_1 -  \frac{N}{4M_2} \cdot M_2 \right)}{N + \frac{N}{4M_2}} \\
    &\overset{(a)}{\ge} \frac{\frac{N}{16M_2} \cdot \frac{N}{2}}{ N + \frac{N}{12} } \\
    &\ge \frac{N}{35M_2} \\
    &\ge \frac{1}{35} \min \left\{ K_1K_2, \frac{N}{M_2} \right\} . 
\end{align*}
where $(a)$ follows from
\begin{equation*}
    \frac{N}{4M_2} \le \frac{K_2}{3} = \frac{K_1K_2}{3K_1} \le \frac{N}{3K_1} \le \frac{N}{12}
\end{equation*}
using $N \ge K_1K_2$ and $K_1 \ge 4$.   
Combining with \eqref{Eqn:RateR1Reg1}, we have 
\begin{align}
    \label{Eqn:BGRegimeICaseB}
    R_1^{\textup{lb}}(M_1, M_2) &\ge \frac{1}{35}R_1 (\alpha^\star, \beta^\star) .
\end{align}

\bigskip

\subsubsection{$\displaystyle 0 \le M_1 < \frac{N}{4}, \ \ \frac{N}{4} \le M_2 \le N$}
\label{Sec:RegimeICaseC}

We trivially have  
\begin{align*}
    R_1^{\textup{lb}}(M_1, M_2) 
    \ge 0 
    \ge \frac{N}{M_2} - 4 
    \ge \min\left\{ K_1K_2, \frac{N}{M_2} \right\}  - 4. 
\end{align*}
Combined with \eqref{Eqn:RateR1Reg1}, this yields 
\begin{align}
    \label{Eqn:BGRegimeICaseC}
    R_1^{\textup{lb}}(M_1, M_2) \ge R_1(\alpha^\star, \beta^\star) - 4. \\
    \nonumber
\end{align}

Sections~\ref{Sec:RegimeICaseA},~\ref{Sec:RegimeICaseB},
and~\ref{Sec:RegimeICaseC} cover all the cases in regime I.  Combining
\eqref{Eqn:BGRegimeICaseA}, \eqref{Eqn:BGRegimeICaseB}, and
\eqref{Eqn:BGRegimeICaseC}, it follows that the  achievable rate $R_1
(\alpha^\star, \beta^\star)$ and the lower bound $R_1^{\textup{lb}}(M_1,
M_2)$ are within a constant multiplicative and additive gap for this
regime. 

\subsection*{\textup{Regime II:} $\displaystyle M_1 + M_2K_2 < N$} 
\addtocounter{subsection}{1}
\addtocounter{subsubsection}{-3}
\label{Sec:BGRegimeII}

For this regime, recall from \eqref{Eqn:UpperBoundsR1} that the
achievable rate $R_1(\alpha^\star , \beta^\star)$  is upper bounded as 
\begin{equation}
    \label{Eqn:RegimeIIR1}
    R_1(\alpha^\star , \beta^\star)  
    \le \min \left\{ K_1K_2, \frac{M_1}{M_1 + M_2K_2} 
    \cdot \frac{(N - M_1)K_2}{M_1 + M_2K_2} + 
    \frac{M_2K_2}{M_1 + M_2K_2} \cdot \frac{NK_2 - M_1}{M_1 + M_2K_2} \right\} .
\end{equation}
For characterizing the gap between the achievable rate and the lower
bounds, we further divide this regime into the following subregimes: 
\begin{alignat*}{2}
    &\hspace{-2in}\textup{\ref{Sec:RegimeIICaseA})}& \ \displaystyle 0 \le M_1 < \frac{N}{K_1}&, \ \ 0 \le M_2 < \frac{N}{K_1K_2}, \\[1em]
    &\hspace{-2in}\textup{\ref{Sec:RegimeIICaseB})}& \ \displaystyle 0  \le M_1 < \frac{N}{K_1}&, \ \ \frac{N}{K_1K_2} \le M_2 < \frac{N}{3K_2},\\[1em]
    &\hspace{-2in}\textup{\ref{Sec:RegimeIICaseC})}& \ \displaystyle 0 \le M_1 < \frac{N}{K_1}&, \ \ \frac{N}{3K_2} \le M_2 < \frac{N}{4},\\[1em]
    &\hspace{-2in}\textup{\ref{Sec:RegimeIICaseD})}& \ \displaystyle \frac{N}{K_1}\le M_1 < \frac{N}{4}&, \ \   0 \le M_2 < \frac{N}{4K_2},\\[1em]
    &\hspace{-2in}\textup{\ref{Sec:RegimeIICaseE})}& \ \displaystyle  \frac{N}{K_1}\le M_1 < \frac{N}{4}&, \ \frac{N}{4K_2} \le M_2 < \frac{N}{4},\\[1em]
    &\hspace{-2in}\textup{\ref{Sec:RegimeIICaseF})}& \ \displaystyle  \frac{N}{4} \le M_1 \le N&, \ \  0 \le M_2 < \frac{N - M_1}{2K_2},\\[1em]
    &\hspace{-2in}\textup{\ref{Sec:RegimeIICaseG})}& \ \displaystyle   \frac{N}{4} \le M_1 \le N&, \ \  \frac{N - M_1}{2K_2} \le M_2 < \frac{N - M_1}{K_2}.  
\end{alignat*}
The subregimes above only consider $M_2 < N / 4$ since from the definition of regime~II, we have 
\begin{equation*}
    M_2 < \frac{N - M_1}{K_2}  \le \frac{N}{K_2} \le \frac{N}{4}
\end{equation*}
using $K_2 \ge 4$. We now consider the different subregimes one by one.

\bigskip

\subsubsection{ $\displaystyle 0 \le M_1 < \frac{N}{K_1}, \ \ 0 \le M_2 < \frac{N}{K_1K_2}$ } 
\label{Sec:RegimeIICaseA}

Let  
\begin{align*}
    s_1 & = \left\lfloor \frac{K_1}{3} \right\rfloor, \\
    s_2 & = K_2
\end{align*}
in the lower bound \eqref{Eqn:LowerBoundR1App2}. This is a valid choice
since $K_1 \ge 4$, and thus $\left\lfloor K_1 / 3 \right\rfloor \ge 1$. Evaluating
\eqref{Eqn:LowerBoundR1App2},  we obtain 
\begin{align*}
    R_1^{\textup{lb}}\left(M_1, M_2\right)  &\ge \frac{\left\lfloor \frac{K_1}{3} \right\rfloor K_2 \left( N - \left\lfloor \frac{K_1}{3} \right\rfloor M_1 - \left\lfloor \frac{K_1}{3} \right\rfloor K_2 M_2 \right)}{ N +  \left\lfloor \frac{K_1}{3} \right\rfloor K_2} \\
    &\overset{(a)}{\ge} \frac{ \frac{K_1K_2}{6}  \left( N - \frac{M_1K_1}{3} - \frac{M_2K_1K_2}{3} \right)}{ N +  \frac{K_1K_2}{3}} \\
    &\overset{(b)}{\ge} \frac{ \frac{K_1K_2}{6}  \left( N - \frac{N}{3} - \frac{N}{3} \right)}{ N +  \frac{N}{3}} \\
    &=  \frac{K_1K_2}{24} \\
    &\ge \frac{1}{24} \min \left\{ K_1K_2, \frac{M_1}{M_1 + M_2K_2} \cdot \frac{(N - M_1)K_2}{M_1 + M_2K_2} + \frac{M_2K_2}{M_1 + M_2K_2} \cdot \frac{NK_2 - M_1}{M_1 + M_2K_2} \right\}
\end{align*}
where $(a)$ follows since $\lfloor x \rfloor \ge x / 2$ for any $x \ge
1$; and  $(b)$ follows from $M_1 < N /K_1$, $M_2 < N / (K_1K_2)$,
and $N \ge K_1K_2$. Combining with \eqref{Eqn:RegimeIIR1}, we have 
\begin{equation}
    \label{Eqn:BGRegimeIICaseA}
    R_1^{\textup{lb}}\left(M_1, M_2\right) 
    \ge \frac{1}{24} R_1(\alpha^\star, \beta^\star) .
\end{equation}

\bigskip

\subsubsection{$\displaystyle 0 \le M_1 < \frac{N}{K_1}, \ \ \frac{N}{K_1K_2} \le M_2 < \frac{N}{3K_2}$}
\label{Sec:RegimeIICaseB}

Let 
\begin{align*}
    s_1 & = \left\lfloor \frac{N}{3M_2K_2} \right\rfloor, \\
    s_2 & = K_2  
\end{align*}
in \eqref{Eqn:LowerBoundR1App2}. Note that this is a valid choice since 
\begin{equation*}
    1 \le  \left\lfloor \frac{N}{3M_2K_2} \right\rfloor \le \frac{N}{3M_2K_2} \le \frac{K_1}{3} .
\end{equation*}
Substituting $s_1, s_2$ in \eqref{Eqn:LowerBoundR1App2}, we have 
\begin{align*}
    R_1^{\textup{lb}}\left(M_1, M_2\right)  &\ge \frac{ \left\lfloor \frac{N}{3M_2K_2} \right\rfloor  K_2\left(N - \left\lfloor \frac{N}{3M_2K_2} \right\rfloor M_1 -  \left\lfloor \frac{N}{3M_2K_2} \right\rfloor K_2 M_2 \right)}{ N + \left\lfloor \frac{N}{3M_2K_2} \right\rfloor K_2 } \\
    &\overset{(a)}{\ge} \frac{ \frac{N}{6M_2} \left( N - \frac{NM_1}{3M_2K_2} - \frac{N }{3}\right) }{ N + \frac{N}{3M_2} } \\
    &= \frac{ \frac{N}{6M_2} \left( \frac{2}{3} - \frac{M_1}{3M_2K_2} \right)}{ 1 + \frac{1}{3M_2} }\\
        &\overset{(b)}{\ge} \frac{ \frac{N}{6M_2} \left( \frac{2}{3} - \frac{1}{3} \right)}{ 1 + \frac{1}{3} }\\ 
    &\ge \frac{N}{24M_2} \\
    &\ge \frac{1}{24} \min \left\{ K_1K_2, \frac{N}{M_2} \right\} \\
    &\ge \frac{1}{24} \min \left\{ K_1K_2, \frac{M_1}{M_1 + M_2K_2} \cdot \frac{(N - M_1)K_2}{M_1 + M_2K_2} + \frac{M_2K_2}{M_1 + M_2K_2} \cdot \frac{NK_2 - M_1}{M_1 + M_2K_2} \right\}
\end{align*}
where $(a)$ follows from $\lfloor x \rfloor \ge x / 2$ for any $x \ge 1$; and 
 $(b)$ follows from
$M_1 < N /K_1$, $M_2 \ge N / (K_1K_2)$, and since 
\begin{equation*}
    \frac{1}{3M_2} \le \frac{K_1K_2}{3N} \le \frac{1}{3} 
\end{equation*}
using $N \ge K_1K_2$. Combined with
\eqref{Eqn:RegimeIIR1}, we have 
\begin{equation}
    \label{Eqn:BGRegimeIICaseB}
    R_1^{\textup{lb}}\left(M_1, M_2\right) \ge \frac{1}{24} R_1(\alpha^\star, \beta^\star) .
\end{equation}

\bigskip

\subsubsection{$\displaystyle 0 \le M_1 < \frac{N}{K_1}, \ \ \frac{N}{3K_2} \le M_2 < \frac{N}{4}$}
\label{Sec:RegimeIICaseC}

Let 
\begin{align*}
    s_1 & = 1 , \\
    s_2 & = \left\lfloor \frac{N}{4M_2} \right\rfloor  
\end{align*}
in the lower bound in \eqref{Eqn:LowerBoundR1App2}. This is a
valid choice since 
\begin{equation}
    \label{Eqn:Temp2}
    1 = \left\lfloor \frac{N}{4 \cdot N / 4} \right\rfloor \le   \left\lfloor \frac{N}{4M_2} \right\rfloor \le  \frac{N}{4M_2} \le \frac{3K_2}{4} .
\end{equation}
Evaluating \eqref{Eqn:LowerBoundR1App2}, we obtain 
\begin{align*}
    R_1^{\textup{lb}}\left(M_1, M_2\right)  &\ge \frac{ \left\lfloor \frac{N}{4M_2} \right\rfloor  \left( N - M_1 - \left\lfloor \frac{N}{4M_2} \right\rfloor M_2\right) }{ N + \left\lfloor \frac{N}{4M_2} \right\rfloor} \\
    &\overset{(a)}{\ge} \frac{ \frac{N}{8M_2} \left( N - M_1 - \frac{N }{4} \right)}{ N + \frac{N}{4M_2} }  \\
    &=  \frac{ \frac{N}{8M_2} \left( \frac{3}{4} -  \frac{M_1}{N} \right)}{1 + \frac{1}{4M_2} } \\
    &\overset{(b)}{\ge} \frac{ \frac{N}{8M_2} \left( \frac{3}{4} - \frac{1}{4} \right)}{1 + \frac{3}{16} } \\ 
    &\ge \frac{N}{19M_2} \\
    &\ge \frac{1}{19} \min \left\{ K_1K_2, \frac{M_1}{M_1 + M_2K_2} \cdot \frac{(N - M_1)K_2}{M_1 + M_2K_2} + \frac{M_2K_2}{M_1 + M_2K_2} \cdot \frac{NK_2 - M_1}{M_1 + M_2K_2} \right\}
\end{align*}
where $(a)$ follows from $\lfloor x \rfloor \ge x / 2$ for any $x \ge 1$; and $(b)$
follows since $M_1 < N /K_1\le N / 4$ using $K_1 \ge 4$, and 
\begin{equation*}
    \frac{1}{4M_2} \le \frac{3K_2}{4N} = \frac{3K_1K_2}{4K_1N} \le \frac{3}{16}
\end{equation*}
using \eqref{Eqn:Temp2}, $N \ge K_1K_2$, and $K_1 \ge 4$. Combining with \eqref{Eqn:RateR1Reg1}, we
have 
\begin{equation}
    \label{Eqn:BGRegimeIICaseC}
    R_1^{\textup{lb}}\left(M_1, M_2\right) \ge \frac{1}{19} R_1(\alpha^\star, \beta^\star).
\end{equation}

\bigskip

\subsubsection{$\displaystyle { \frac{N}{K_1}\le M_1 < \frac{N}{4}, \ 0 \le M_2 < \frac{N}{4K_2} }$}
\label{Sec:RegimeIICaseD}

Let
\begin{align*}
    s_1 & = \left\lfloor \frac{N}{2(M_1 + M_2K_2)}  \right\rfloor, \\
    s_2 & = K_2 
\end{align*}
in \eqref{Eqn:LowerBoundR1App2}. Note that this is a valid choice since  
\begin{align}
\label{Eqn:IneqConst} 
    1 =  \left\lfloor  \frac{N}{2( N / 4 + N/ 4)}   \right\rfloor   \le &\left\lfloor \frac{N}{2(M_1 + M_2K_2)} \right\rfloor 
    \le \frac{N}{2(M_1 + M_2K_2)} \le \frac{N}{2M_1} \le \frac{K_1}{2}.
\end{align}
Substituting $s_1, s_2$ in \eqref{Eqn:LowerBoundR1App2}, we obtain 
\begin{align}
    \nonumber
    R_1^{\textup{lb}}(M_1, M_2) &\ge \frac{ \left\lfloor \frac{N}{2(M_1 + M_2K_2)} \right\rfloor K_2 \left( N -  \left\lfloor \frac{N}{2(M_1 + M_2K_2)} \right\rfloor (M_1 + M_2K_2 )\right) }{ N + \left\lfloor \frac{N}{2(M_1 + M_2K_2)} \right\rfloor  K_2 } \\
    \nonumber
    &\overset{(a)}{\ge} \frac{ \frac{N}{4(M_1 + M_2K_2)} K_2 \left( N - \frac{N}{2(M_1 + M_2K_2)} (M_1 + M_2K_2 ) \right) }{ N + \frac{NK_2}{2M_1} } \\
    \nonumber
    &\overset{(b)}{\ge}  \frac{ \frac{NK_2}{4(M_1 + M_2K_2)} \left( N - \frac{N}{2}\right) }{ N + \frac{N}{2} } \\
    \nonumber
    &=  \frac{NK_2}{12(M_1 + M_2K_2)} \\
    \nonumber
    &\ge \frac{1}{12} \min \left\{ K_1K_2, \frac{NK_2}{M_1 + M_2K_2} \right\} \\
    &\ge \frac{1}{12} \min \left\{ K_1K_2, \frac{M_1}{M_1 + M_2K_2} \cdot \frac{(N - M_1)K_2}{M_1 + M_2K_2} + \frac{M_2K_2}{M_1 + M_2K_2} \cdot \frac{NK_2 - M_1}{M_1 + M_2K_2} \right\}
    \nonumber
\end{align}
where $(a)$ follows since $\lfloor x \rfloor \ge x / 2$ for any $x \ge
1$; and $(b)$ follows since
\begin{equation*}
    \frac{NK_2}{2M_1} \le \frac{K_1K_2}{2} \le \frac{N}{2} 
\end{equation*}
using \eqref{Eqn:IneqConst} and $N \ge K_1K_2$. 
Combining with \eqref{Eqn:RegimeIIR1}, we have 
\begin{align}
    R_1^{\textup{lb}}(M_1, M_2) &\ge \frac{1}{12} R_1(\alpha^\star, \beta^\star) . 
    \label{Eqn:BGRegimeIICaseD}
\end{align}

\bigskip

\subsubsection{$ {\displaystyle  \frac{N}{K_1}\le M_1 < \frac{N}{4}, \ \frac{N}{4K_2} \le M_2 < \frac{N}{4} }$}   
\label{Sec:RegimeIICaseE}

Let 
\begin{align*}
    (s_1 , s_2)  &= 
    \begin{cases}
        \left( \left\lfloor\frac{N}{4M_1} \right\rfloor, \left\lfloor \frac{M_1}{M_2} \right\rfloor \right) &\mbox{ if } M_1 \ge M_2, \\[.75em]
        \left( \left\lfloor \frac{N}{4M_2}\right\rfloor, 1 \right) &\mbox{ otherwise,}  
    \end{cases}
\end{align*}
in \eqref{Eqn:LowerBoundR1App2}. This is a valid choice since for $M_1 \ge M_2$, we have
\begin{align*}
    1 =      \left\lfloor\frac{N}{4 \cdot N / 4} \right\rfloor \le &\left\lfloor\frac{N}{4M_1} \right\rfloor \le \frac{N}{4M_1} \le \frac{K_1}{4} ,\\[.75em]
    1 \le  &\left\lfloor \frac{M_1}{M_2} \right\rfloor \le \frac{M_1}{M_2}  \le \frac{N/ 4}{N / (4K_2)} = K_2,
\end{align*}
and for $M_1 < M_2$, we have 
\begin{align*}
    1 =      \left\lfloor\frac{N}{4 \cdot N / 4} \right\rfloor \le &\left\lfloor\frac{N}{4M_2} \right\rfloor \le \left\lfloor\frac{N}{4M_1} \right\rfloor \le \frac{N}{4M_1} \le \frac{K_1}{4}.
\end{align*}
Note that $s_1 \le N / (4M_1)$ and $s_1s_2 \le N / (4M_2)$. Further, since $\lfloor x \rfloor \ge x / 2$ for any $x \ge 1$, we have $s_1s_2 \ge N / (16M_2)$. Also, note that  
\begin{equation*}
    \frac{N}{4M_2} \le  K_2 = \frac{K_1K_2}{K_1} \le \frac{K_1K_2}{4} \le \frac{N}{4}, 
\end{equation*}
using $N \ge K_1K_2$ and $K_1 \ge 4$. Finally, substituting $s_1, s_2$ in \eqref{Eqn:LowerBoundR1App2}, we obtain 
\begin{align}
    \nonumber
    R_{1}^{\textup{lb}}\left(M_1, M_2\right) &\ge \frac{ \frac{N}{16M_2} \left( N - \frac{N}{4M_1} \cdot M_1 - \frac{N}{ 4M_2} \cdot M_2 \right)}{ N + \frac{N}{4M_2} }  \\
    \nonumber
    &\ge \frac{ \frac{N}{16M_2} \left( N - \frac{N}{2} \right)}{ N + \frac{N}{4} } \\
    \nonumber
    &=  \frac{N}{40M_2} \\
    \nonumber
    &\ge \frac{1}{40} \min \left\{ K_1K_2, \frac{M_1}{M_1 + M_2K_2} \cdot \frac{(N - M_1)K_2}{M_1 + M_2K_2} + \frac{M_2K_2}{M_1 + M_2K_2} \cdot \frac{NK_2 - M_1}{M_1 + M_2K_2} \right\}. 
\end{align}
Combining with \eqref{Eqn:RegimeIIR1}, we have 
\begin{align}
    \label{Eqn:BGRegimeIICaseE}
    R_1^{\textup{lb}}(M_1, M_2) &\ge \frac{1}{40} R_1(\alpha^\star, \beta^\star).
\end{align}

\bigskip

\subsubsection{$\displaystyle  \frac{N}{4} \le M_1 \le N, \ \  0 \le M_2 < \frac{N - M_1}{2K_2}$}
\label{Sec:RegimeIICaseF}

Substituting  $s_1 = 1, s_2 = K_2$ in the lower bound \eqref{Eqn:LowerBoundR1App2}, we obtain 
\begin{align}
    \nonumber
    R_{1}^{\textup{lb}}(M_1, M_2) &\ge \frac{K_2\left(N - M_1 - M_2K_2\right)}{N + K_2} \\
    \nonumber
    &\overset{(a)}{\ge} \frac{K_2\left(N - M_1 - (N - M_1)/2\right)}{N + \frac{N}{4}}\\
    &=\frac{2K_2(N- M_1)}{5N} 
    \label{Eqn:LowerBoundRegimeIICaseF}
\end{align}
where $(a)$ follows since 
\begin{equation*}
    K_2 = \frac{K_1K_2}{K_1} \le \frac{K_1K_2}{4} \le \frac{N}{4}
\end{equation*}
using $N \ge K_1K_2$ and $K_1 \ge 4$. On the other hand, from
\eqref{Eqn:RegimeIIR1} we obtain  
\begin{align*}
    \nonumber
    R_{1}(\alpha^\star, \beta^\star) &\le \frac{M_1}{M_1 + M_2K_2} \cdot  \frac{K_2(N- M_1)}{M_1 + M_2K_2} + \frac{M_2K_2}{M_1+M_2K_2} \cdot \frac{NK_2 - M_1 }{M_1 + M_2K_2} \\
    \nonumber
    &\le  \frac{K_2(N- M_1)}{M_1 + M_2K_2} + \frac{M_2K_2}{M_1+M_2K_2} \cdot \frac{NK_2}{M_1} \\
    \nonumber
    &\overset{(a)}{\le}  \frac{K_2(N- M_1)}{M_1 + M_2K_2} + \frac{(N - M_1)}{2} \cdot \frac{N}{M_1} \cdot \frac{K_2}{(M_1 + M_2K_2)} \\
    \nonumber
    &= \frac{K_2(N- M_1)}{M_1 + M_2K_2}  \left( 1 + \frac{N}{2M_1} \right) \\
    &\overset{(b)}{\le} \frac{3K_2(N- M_1)}{M_1 + M_2K_2} 
\end{align*}
where $(a)$ follows since $M_2 < (N - M_1) / (2K_2)$ for this case; and
$(b)$ follows since $M_1 \ge N / 4$. Combining with
\eqref{Eqn:LowerBoundRegimeIICaseF}, we obtain
\begin{align}
    \nonumber
    R_{1}^{\textup{lb}}(M_1, M_2) &\ge \frac{2K_2(N- M_1)}{5N}  \\
    \nonumber
    &=\frac{2}{5 \cdot 3} \cdot \frac{M_1 + M_2K_2}{N} \cdot \frac{3K_2(N - M_1)}{M_1 + M_2K_2} \\
    \nonumber
    &\overset{(a)}{\ge} \frac{2}{15} \cdot \frac{N / 4}{N} \cdot \frac{3K_2(N - M_1)}{M_1 + M_2K_2} \\
    \label{Eqn:BGRegimeIICaseF}
    &\ge \frac{1}{30} R_{1}(\alpha^\star, \beta^\star)
\end{align}
where $(a)$ follow since $M_1 \ge N / 4$. 

\bigskip

\subsubsection{$\displaystyle  \frac{N}{4} \le M_1 \le N, \ \  \frac{N - M_1}{2K_2} \le M_2 < \frac{N - M_1}{K_2}$ }
\label{Sec:RegimeIICaseG}

Let 
\begin{align*}
    s_1 & = 1, \\
    s_2 & = \left\lfloor\frac{N - M_1}{2M_2}\right\rfloor
\end{align*}
in \eqref{Eqn:LowerBoundR1App2}. This is a valid choice since $K_2 \ge
4$ and $M_1 + M_2K_2 < N$, so that 
\begin{equation*}
    1 \le \left\lfloor\frac{N - M_1}{K_2M_2}\right\rfloor  
    \le \left\lfloor\frac{N - M_1}{2M_2}\right\rfloor 
    \le \frac{N - M_1}{2M_2} 
    \le K_2 . 
\end{equation*}
Substituting $s_1, s_2$ in \eqref{Eqn:LowerBoundR1App2}, we obtain
\begin{align}
    \nonumber
    R_{1}^{\textup{lb}}(M_1, M_2) &\ge \left\lfloor\frac{N - M_1}{2M_2}\right\rfloor \frac{\left(N - M_1 - \left\lfloor\frac{N - M_1}{2M_2}\right\rfloor M_2\right)}{N + \left\lfloor\frac{N - M_1}{2M_2}\right\rfloor} \\
    \nonumber
    &\overset{(a)}{\ge} \frac{N - M_1}{4M_2} \frac{\left(N - M_1 - \frac{N - M_1}{2M_2} M_2 \right)}{N + \frac{N}{4}} \\
    \nonumber
    &= \frac{N - M_1}{4M_2} \cdot \frac{2(N - M_1)}{5N} \\
    &= \frac{(N - M_1)^2}{10M_2N} 
    \label{Eqn:LowerBoundRegimeIICaseG}
\end{align}
where $(a)$ follows from $\lfloor x \rfloor \ge x / 2$ for any $x \ge 1$, and 
\begin{equation*}
    \left\lfloor \frac{N - M_1}{2M_2}\right\rfloor \le \frac{N - M_1}{2M_2} \le K_2 \le \frac{N}{K_1} \le \frac{N}{4}
\end{equation*}
using $N \ge K_1K_2$ and $K_1 \ge 4$. On the other hand, from
\eqref{Eqn:RegimeIIR1} we obtain
\begin{align*}
    R_{1}(\alpha^\star, \beta^\star) &\le \frac{M_1}{M_1 + M_2K_2} \cdot \frac{K_2(N- M_1)}{M_1 + M_2K_2} + \frac{M_2K_2}{M_1+M_2K_2} \cdot \frac{NK_2 - M_1 }{M_1 + M_2K_2} \\
    &\overset{(a)}{\le} \frac{K_2(N- M_1)}{M_1 + M_2K_2} + \frac{N - M_1}{M_1 + M_2K_2} \cdot \frac{NK_2}{M_1 + M_2K_2} \\
    &=  \frac{K_2(N- M_1)}{M_1 + M_2K_2} \left(1  + \frac{N}{M_1 + M_2K_2} \right) \\
    &\overset{(b)}{\le} \frac{3K_2(N- M_1)}{M_1 + M_2K_2} 
\end{align*}
where $(a)$ follows since $M_1 + M_2K_2 < N$  for regime~II and $(b)$
follows since $M_1 + M_2K_2 \ge M_1 + (N - M_1)/2 \ge N / 2$ for this
case. Combining with \eqref{Eqn:LowerBoundRegimeIICaseG}, we have
\begin{align}
    \nonumber
    R_{1}^{\textup{lb}}(M_1, M_2)  &\ge  \frac{( N- M_1)^2}{10M_2N} \\
    \nonumber
    &= \frac{1}{30} \cdot \frac{3K_2(N-M_1)}{M_1 + M_2K_2}  \cdot \frac{N - M_1}{M_2K_2} \cdot \frac{M_1 + M_2K_2} {N}  \\
    \nonumber
    &\overset{(a)}{\ge} \frac{1}{30} \cdot \frac{3K_2(N-M_1)}{M_1 + M_2K_2}  \cdot  1 \cdot \frac{1}{2}\\
    \label{Eqn:BGRegimeIICaseG}
    &\ge \frac{1}{60} R_{1}(\alpha^\star, \beta^\star)
\end{align}
where $(a)$ follows since $N / 2 \le M_1 + M_2K_2 < N$ for this case. 

\bigskip

Sections~\ref{Sec:RegimeIICaseA} - \ref{Sec:RegimeIICaseG} cover all the
cases in regime~II.  Combining \eqref{Eqn:BGRegimeIICaseA},
\eqref{Eqn:BGRegimeIICaseB}, \eqref{Eqn:BGRegimeIICaseC},
\eqref{Eqn:BGRegimeIICaseD}, \eqref{Eqn:BGRegimeIICaseE},
\eqref{Eqn:BGRegimeIICaseF}, and \eqref{Eqn:BGRegimeIICaseG} shows that
the  achievable rate $R_1 (\alpha^\star, \beta^\star)$ and the lower
bound $R_1^{\textup{lb}}(M_1, M_2)$ are within a constant multiplicative
and additive gap in this regime.

\subsection*{\textup{Regime III:} $\displaystyle M_1 + M_2K_2 \ge N, \ \frac{N}{4} \le M_1 \le N$} 
\addtocounter{subsection}{1}
\addtocounter{subsubsection}{-7}
\label{Sec:BGRegimeIII}

For this regime, recall from \eqref{Eqn:UpperBoundsR1} that the
achievable rate $R_1(\alpha^\star, \beta^\star)$ is upper bounded as
\begin{equation}
    \label{Eqn:RegimeIIIR1}
    R_1(\alpha^\star , \beta^\star)  \le \frac{4(N - M_1)^2}{3NM_2} .
\end{equation}
To characterize the gap between the achievable rate and the lower
bounds, we further divide regime~III into the two subregimes \\[1em]
\ref{Sec:RegimeIIICaseA}) 
$\displaystyle \frac{N}{4} \le M_1 \le N, \  \frac{N - M_1}{K_2} \le M_2 <  \frac{N - M_1}{2}$,\\[1em]
\ref{Sec:RegimeIIICaseB}) 
$\displaystyle  \frac{N}{4} \le M_1 \le N, \  \frac{N - M_1}{2} \le M_2 \le N$.\\[1em]
We now consider the subregimes one by one.

\bigskip

\subsubsection{$\displaystyle \frac{N}{4} \le M_1 \le N, \ \frac{N - M_1}{K_2} \le M_2 <  \frac{N - M_1}{2}$} 
\label{Sec:RegimeIIICaseA}

Let 
\begin{align*}
    s_1 & = 1, \\
    s_2 & = \left\lfloor\frac{N - M_1}{2M_2}\right\rfloor
\end{align*}
in the lower bound \eqref{Eqn:LowerBoundR1App2}. This is a valid choice
since 
\begin{equation}
    \label{Eqn:TempRegimeIIICaseA}
    1 \le   \left\lfloor\frac{N - M_1}{2M_2}\right\rfloor  
    \le \frac{N - M_1}{2M_2} 
    \le \frac{K_2}{2}.
\end{equation}
Substituting $s_1, s_2$ in \eqref{Eqn:LowerBoundR1App2}, we obtain
\begin{align*}
    R_{1}^{\textup{lb}}(M_1, M_2) &\ge \left\lfloor\frac{N - M_1}{2M_2}\right\rfloor \frac{\left(N - M_1 - \left\lfloor\frac{N - M_1}{2M_2}\right\rfloor M_2\right)}{N + \left\lfloor\frac{N - M_1}{2M_2}\right\rfloor} \\
    &\overset{(a)}{\ge} \frac{N - M_1}{4M_2} \frac{\left(N - M_1 - \frac{N - M_1}{2M_2} M_2 \right)}{N + \frac{N}{8}} \\
    &= \frac{(N - M_1)^2}{9M_2N} 
\end{align*}
where $(a)$ follows from $\lfloor x \rfloor \ge x / 2$ for any $x \ge 1$ and 
\begin{equation*}
    \left\lfloor\frac{N - M_1}{2M_2}\right\rfloor 
    \le \frac{K_2}{2} 
    = \frac{K_1K_2}{2K_1} 
    \le \frac{N}{8}. 
\end{equation*}
using \eqref{Eqn:TempRegimeIIICaseA}, $N \ge K_1K_2$, and $K_1 \ge 4$.
Combining with \eqref{Eqn:RegimeIIIR1}, we have
\begin{align}
    \label{Eqn:BGRegimeIIICaseA}
    R_{1}^{\textup{lb}}(M_1, M_2) &\ge\frac{( N- M_1)^2}{9M_2N} = \frac{1}{12} \cdot \frac{ 4(N- M_1)^2}{3NM_2}  \ge \frac{1}{12} R_1(\alpha^\star, \beta^\star) .
\end{align}

\bigskip

\subsubsection{$\displaystyle  \frac{N}{4} \le M_1 \le N, \  \frac{N - M_1}{2} \le M_2 \le N$}  
\label{Sec:RegimeIIICaseB}
We trivially have
\begin{align*}
    R_1^{\textup{lb}}(M_1, M_2) \ge 0 = \frac{8}{3} - \frac{8}{3}. 
\end{align*}
Combining with \eqref{Eqn:RegimeIIIR1} and using $(N - M_1) / M_2 \le 2$
for this case, we have 
\begin{align}
    \nonumber
    R_1^{\textup{lb}}(M_1, M_2) &\ge \frac{4}{3} \cdot 2 \cdot 1 - \frac{8}{3} \\
    \nonumber
    &\ge \frac{4}{3} \cdot \frac{N - M_1}{M_2} \cdot \frac{N - M_1}{N} - \frac{8}{3} \\ 
    \nonumber
    &= \frac{ 4(N- M_1)^2}{3NM_2} - \frac{8}{3} \\
    \label{Eqn:BGRegimeIIICaseB}
    &\ge R_1(\alpha^\star, \beta^\star) - \frac{8}{3}. \\
    \nonumber
\end{align}
Sections~\ref{Sec:RegimeIIICaseA} and ~\ref{Sec:RegimeIIICaseB} cover
all the cases in regime III.  Combining \eqref{Eqn:BGRegimeIIICaseA} and
\eqref{Eqn:BGRegimeIIICaseB}, it follows that the  achievable rate $R_1
(\alpha^\star, \beta^\star)$ and the lower bound $R_1^{\textup{lb}}(M_1,
M_2)$ are within a constant multiplicative and additive gap for
regime~III. 

Regimes~I, II, and III cover all possible values for $(M_1, M_2)$. For
each regime, we have shown that the achievable rate $R_1$ for the
generalized caching scheme is within a constant additive and
multiplicative gap of the minimum feasible rate. In particular, for any
$M_1, M_2$, and any feasible rate pair $(R_1, R_2) \in \mc{R}^\star(M_1,
M_2)$, we have 
\begin{equation*}
    R_1 \ge R_1^{\textup{lb}}(M_1, M_2) \ge \frac{1}{60} R_1 (\alpha^\star, \beta^\star) - 4. 
\end{equation*}

\section{Gap between achievable rate $R_2(\alpha^\star, \beta^\star)$ and lower bound $R_{2}^{\textup{lb}}\left(M_1, M_2\right)$}
\label{Sec:GapR2}

We prove that the rate $R_2(\alpha^\star, \beta^\star)$ for the
generalized caching scheme, as described in \eqref{Eqn:UpperBoundsR2},
is within a constant additive and multiplicative gap of  the
corresponding lower bound $R_2^{\textup{lb}}(M_1, M_2)$, as derived in
Appendix~\ref{Sec:LowerBounds}, for all values of $M_1, M_2$. As before,
we focus on the case where $K_1, K_2 \ge 4$. The case of $K_1 < 4$ or
$K_2 < 4$ is easily analyzed using the results of \cite{CachingUM}. 

From \eqref{Eqn:UpperBoundsR2}, we have the following upper bound on the
achievable rate $R_2(\alpha^\star , \beta^\star)$ of the generalized
caching scheme for any $M_1, M_2$:
\begin{equation}
    \label{Eqn:R2UB}
    R_2(\alpha^\star , \beta^\star) \le 4 \min \left\{ K_2, \frac{N}{M_2} \right\} .
\end{equation}
On the other hand, recall the following lower bound on the rate $R_2$
from \eqref{Eqn:LowerBoundR2}: 
\begin{align}
    R_{2}^{\textup{lb}}\left(M_1, M_2\right) 
    & = \max_{t \in \{1,2\ldots, K_2\} }   \ \frac{t(N - tM_2)}{ N + t} . 
    \label{Eqn:LowerBoundR2App}
\end{align}
To characterize the gap between the achievable rate and the lower
bounds, we study two different cases. 

\begin{enumerate}
    \item $\displaystyle 0 \le M_2 < \frac{N}{4}$, \\
    \item $\displaystyle \frac{N}{4} \le M_2 \le N$. \\
\end{enumerate}

We now consider the two cases one by one.

\bigskip

\begin{enumerate}
    \item $0 \le M_2 < \displaystyle \frac{N}{4}$: Let
        \begin{equation*}
            t = \left\lfloor  \frac{1}{2}\min\left\{K_2,  \frac{N}{M_2} \right\} \right\rfloor 
        \end{equation*}
        in  \eqref{Eqn:LowerBoundR2App}. This is a valid choice since $K_2 \ge 4$, and thus
        \begin{equation*}
            1 \le \left\lfloor \frac{1}{2}\min\left\{K_2, \frac{N}{M_2} \right\} \right\rfloor 
            \le \frac{K_2}{2} .
        \end{equation*}
        Substituting $t$ in  \eqref{Eqn:LowerBoundR2App} yields
        \begin{align}
            \nonumber
            R_2^{\textup{lb}}(M_1, M_2) 
            &\ge  \frac{ \left\lfloor \frac{1}{2}\min\left\{K_2,  \frac{N}{M_2} \right\}  \right\rfloor \left(N - 
            \left\lfloor  \frac{1}{2}\min\left\{K_2,  \frac{N}{M_2} \right\} \right\rfloor M_2\right)}{N +  \left\lfloor \frac{1}{2}
            \min\left\{K_2, \frac{N}{M_2} \right\} \right\rfloor} \\
            \nonumber
            &\overset{(a)}{\ge}   \frac{\frac{1}{4}\min\left\{K_2,  \frac{N}{M_2} \right\}  \left(N - 
             \frac{N}{2} \right)}{N + \frac{N}{8}} \\
            &= \frac{1}{9}\min\left\{K_2, \frac{N}{M_2}  \right\} 
            \label{Eqn:Regime1R2LB}
        \end{align}
        where $(a)$ follows since $\lfloor x \rfloor \ge x / 2$ for any $x \ge 1$ and 
        \begin{equation*}
            \left\lfloor  \frac{1}{2}\min\left\{K_2,  \frac{N}{M_2} \right\} \right\rfloor 
            \le \frac{K_2}{2} = \frac{K_1K_2}{2K_1} \le \frac{N}{8}
        \end{equation*}
        using $N \ge K_1K_2$ and $K_1 \ge 4$. Comparing \eqref{Eqn:R2UB}
        and \eqref{Eqn:Regime1R2LB}, we have 
        \begin{align}
            \label{Eqn:BGR2Case1}
            R_{2}^{\textup{lb}}\left(M_1, M_2\right) 
            &\ge \frac{1}{9}\min\left\{K_2, \frac{N}{M_2}  \right\} 
            = \frac{1}{9 \cdot 4} \cdot 4 \min\left\{K_2,  \frac{N}{M_2}  \right\}  
            \ge \frac{1}{36} R_2(\alpha^\star , \beta^\star) . \\
            \nonumber
        \end{align}
    \item $\displaystyle \frac{N}{4} \le M_2 \le N$: We trivially have  
        \begin{equation*}
            R_{2}^{\textup{lb}}\left(M_1, M_2\right) \ge 0 = 4 \frac{N}{M_2} - 4 \frac{N}{M_2}.
        \end{equation*}
        Combining with \eqref{Eqn:R2UB}, we have
        \begin{align}
            \label{Eqn:BGR2Case2}
            R_{2}^{\textup{lb}}\left(M_1, M_2\right) 
            &\ge 4 \frac{N}{M_2} - 4 \frac{N}{M_2} 
            \ge 4 \min\left\{K_2,  \frac{N}{M_2}  \right\} -  4 \frac{N}{M_2} 
            \ge R_2(\alpha^\star , \beta^\star) - 16 .
        \end{align}
\end{enumerate}

\bigskip

Cases 1) and 2) cover all values of the memory sizes $M_1, M_2$.
Combining \eqref{Eqn:BGR2Case1}, \eqref{Eqn:BGR2Case2}, it follows that
the  achievable rate $R_2 (\alpha^\star, \beta^\star)$ of the
generalized caching scheme and the lower bound $R_2^{\textup{lb}}(M_1,
M_2)$ are within a constant multiplicative and additive gap for all
values of $M_1, M_2$. In particular, for any $M_1, M_2$, and any
feasible rate pair $(R_1, R_2) \in \mc{R}^\star(M_1, M_2)$, we have 
\begin{equation*}
    R_2 \ge R_2^{\textup{lb}}(M_1, M_2) 
    \ge \frac{1}{36} R_2 (\alpha^\star, \beta^\star) - 16. 
\end{equation*}

\end{document}